\documentclass[a4paper,11pt]{article}
 
 \usepackage{amsmath, amssymb, latexsym, amscd, amsthm,amsfonts,amstext}
 \usepackage[mathscr]{eucal}
 \usepackage{graphicx}
 \usepackage{subfig}

 \newtheorem{theorem}{Theorem}[section]

\newtheorem{remark}{Remark}[section]

\title{Imaging of buried objects from experimental backscattering time
dependent measurements using a globally convergent inverse algorithm} 
\author{Nguyen Trung Th\`anh$^{\diamond}$, Larisa Beilina$^{\circ}$, Michael V. Klibanov$^{\diamond}$ \\
and Michael A. Fiddy$^{\circ}$\\
$^\diamond$Department of Mathematics \& Statistics, \\ University of North Carolina at Charlotte, USA\\
  Emails: \texttt{tnguy152@uncc.edu, mklibanv@uncc.edu} \\
$^{\circ}$Department of Mathematical Sciences, \\
Chalmers University of Technology and
Gothenburg University, Sweden \\
Email: \texttt{larisa@chalmers.se}.\\
$^{\circ}$Optoelectronics Center, Univeristy of North Carolina at Charlotte, USA. \\
Email: \texttt{mafiddy@uncc.edu}}
\date{}

\begin{document}
\maketitle

\begin{abstract}
 We consider the problem of imaging of objects buried under the ground using
backscattering experimental time dependent measurements generated by a
single point source or one incident plane wave. In particular, we estimate
dielectric constants of those objects using the globally convergent inverse
algorithm of Beilina and Klibanov. Our algorithm is tested on experimental
data collected using a microwave scattering facility at the University of
North Carolina at Charlotte. There are two main challenges working with this
type of experimental data: (i) there is a huge misfit between these data and
computationally simulated data, and (ii) the signals scattered from the
targets may overlap with and be dominated by the reflection from the
ground's surface. To overcome these two challenges, we propose new data
preprocessing steps to make the experimental data to be approximately the same as the simulated ones, as well as to remove the reflection from the ground's
surface. Results of total 25 data sets of both non blind and blind targets
indicate a good accuracy.
\end{abstract}


\textbf{AMS classification codes:} 35R30, 35L05, 78A46.

\def\bR{\mathbb{R}}
\def\bx{\mathbf{x}}

\section{Introduction}

\label{sec:int}

In \cite{B-K:SISC2008} a globally convergent algorithm for a coefficient
inverse problem (CIP) for a hyperbolic equation was proposed. Since then
several follow-ups on this method were published (see, e.g., \cite%
{B-K:IP2010,B-K:JIIP2012,BK:AA2013,K-F-B-P-S:IP2010}), and results were
summarized in the book \cite{B-K:2012}. In this paper, we demonstrate the
performance of that method for the case of experimentally measured
time-dependent backscattering data for targets buried in a box filled with
dry sand. This mimics the case when the targets are buried under the ground.
Our primary application is in the standoff imaging of explosives, such as,
e.g., land mines and improvised explosive devices (IEDs). A systematic study
of twenty five cases conducted here shows how that method works in
estimating the dielectric constants (equivalently, refractive indices) and
locations of the targets. Since the technique of \cite{B-K:2012} works with
the case when the targets are illuminated by a single point source, we use
here a single location of the source.

This publication is a continuation of three recent works of our group, where
we have treated similar experimental data for targets placed in air \cite%
{BTKF:IP2013,BTKM:IP2014,TBKF:SISC2014}. Compared with the case of targets
in air, there are three main difficulties in imaging of buried targets: (i)
the targets' signals are much weaker than those when the targets are in air,
(ii) the targets' signals may overlap with the reflection from the ground's
surface, which makes it difficult to be distinguished, and (iii) in
particular for the globally convergent method, in which the Laplace
transform of the time-dependent data is used, the reflection from the
ground's surface may dominate the targets' signals after the Laplace
transform, since the kernel of the Laplace transform decays exponentially
with respect to time.

The problem of subsurface imaging can be found in a variety of practical and
engineering applications such as nondestructive testing, landmine and
unexploded ordnance detection, archaeology, remote sensing and medical
imaging, see, e.g., \cite{Pastorino:2010}. For migration-type methods for
estimating geometric information such as shapes, sizes, and locations of
targets, we refer to \cite{Soumekh:1999,Yilmaz:1987}. Our main goal in this
work is to estimate their dielectric constants, which characterize the
targets in terms of their materials. This problem is much more difficult
than estimating the geometric properties. Mathematically speaking, this is a
CIP for the time-dependent wave-like equation in 3-d: we reconstruct a
spatially varying coefficient of that hyperbolic PDE using measurements on
the backscattering part of the boundary of the domain of interest.

The most common method for solving a CIP is the least-squares approach, in
which an objective functional is to be minimized using optimization methods,
see, e.g., \cite{BK:2004,Chavent:2009,E-H-N:1996} and references therein.%
However, it is well known that these objective functionals are non
convex and, as a rule, have multiple local minima and ravines. Therefore,
the convergence of these methods cannot be rigorously guaranteed unless a
good first guess of the exact solution of the CIP is chosen. This means
that they require a priori knowledge of a small neighborhood of the
exact solution, which is not available in many practical situations. 
We call these \emph{locally} convergent methods. It was shown in
section 5.8.4 of \cite{B-K:2012} and in \cite{K-F-B-P-S:IP2010} for
transmitted time-dependent experimental data with a single source that a
locally convergent method starting from the homogeneous background as the
first guess failed, whereas the technique of \cite{B-K:2012} worked well. A
similar conclusion was independently drawn in \cite{Liu} when working with a
different type of experimental data. 

Unlike these, it was rigorously established that the method of \cite%
{B-K:2012} can provide a good approximation of the exact coefficient without
any a priori knowledge of a small neighborhood of this coefficient.
We call a numerical method for a CIP with this property \emph{globally
convergent}.

Since it is extremely hard to develop globally convergent numerical methods
for CIPs, the technique of \cite{B-K:2012} uses a reasonable approximate
mathematical model, see section \ref{sec:gc}. We emphasize that this model
is used only on the first iteration of that method. Due to this model, we
call the method of \cite{B-K:2012} \emph{approximately globally convergent},
or \textit{globally convergent}, in short, see Theorem 2.9.4 in \cite%
{B-K:2012} and Theorem 5.1 in \cite{B-K:JIIP2012} for proofs of the global
convergence in the case when a single location of the point source is used.
We briefly prove here an analog of these theorems for the case when the
point source is replaced by a single incident plane wave. 

Apart from the three difficulties mentioned in the beginning, as we have
pointed out in \cite{TBKF:SISC2014}, the main challenge working with our
experimental data, even when the targets are in air, is a \emph{huge misfit}
between these data and computationally simulated ones. Hence, any inversion
algorithm would fail to produce satisfactory results if being applied to the
raw data. Therefore, the \emph{central procedure} before applying the
globally convergent algorithm to the experimental data is a data
preprocessing procedure, which is inevitably a heuristic one. This procedure
makes the experimental data look somewhat similar to the data provided by
computational simulations. In our opinion, the ultimate justification of
this procedure is the accuracy of reconstruction results, especially
those for \emph{blind} data cases, i.e., for such sets of data for
which correct answers were unknown in advance to the computational team, see
section 5.1 for some details. The preprocessed data are used as the input
for our globally convergent algorithm. A data preprocessing procedure, which
consists of several steps, was proposed in \cite{TBKF:SISC2014}. In
this paper the procedure of \cite{TBKF:SISC2014} is modified to remedy the
aforementioned difficulties related to the case of buried targets. In
particular, we propose a new step for removing the reflection from the
sand's surface. After that, the targets can be treated as being placed in
air, see section \ref{sec:exp}.

We should mention that, unlike dielectric constants, shapes of targets are
more difficult to accurately reconstruct using the globally convergent
method of \cite{B-K:2012}. Only their cross-sections in the measurement
plane can be rather well estimated after the data propagation. To enhance
the reconstruction accuracy of shapes, these results can be used as initial
guesses for locally convergent methods in order to refine the images, see,
e.g., \cite{B-K:IP2010,B-K:2012,BK:AA2013,BTKM:IP2014}. We call the
combination of these two steps a two-stage numerical method. The second step
requires an additional effort and will be reported in a future work. 
We also mention that it was demonstrated in \cite%
{B-K:IP2010,B-K:2012,BK:AA2013} for the transmitted experimental data and in 
\cite{BTKM:IP2014} for the backscattering experimental data that the second
step refines images quite well for targets placed in air.

The rest of the paper is organized as follows. In section \ref{sec:model},
we state the mathematical model and briefly describe the globally convergent
method. In section \ref{sec:3} we outline the main steps of the proof of the
global convergence of the proposed algorithm. Data acquisition and data
preprocessing are described in section \ref{sec:exp}. In section \ref%
{sec:num} we present inversion results of the proposed globally convergent
algorithm for our experimental data. Finally, some conclusions are drawn in
section \ref{sec:con}.


\section{Problem statement and the globally convergent method in brief}

\label{sec:model}

The theory of the globally convergent method of \cite{B-K:SISC2008} using a
single point source has been thoroughly discussed in \cite%
{B-K:2012,B-K:JIIP2012}. The globally convergent method using a plane
incident wave was stated in \cite{TBKF:SISC2014}. We briefly summarize the
main ideas of the latter case here and outline in section \ref{sec:3} its
convergence analysis, which was not included in \cite{TBKF:SISC2014}.

\subsection{Problem statement}

Consider the propagation of the electromagnetic wave in $\mathbb{R}^{3}$
generated by an incident plane wave. Below, $\mathbf{x}=(x,y,z)$ denotes a
point in $\mathbb{R}^{3}$. Since in our experiment only one component $E_{2}(%
\mathbf{x},t)$ of the electric field $E=\left( E_{1},E_{2},E_{3}\right) (%
\mathbf{x},t)$ is generated by the source and the detector measures only
that component of the scattered electric field, we model the wave
propagation by the following Cauchy problem for the scalar wave equation: 
\begin{eqnarray}
&&\epsilon (\mathbf{x})u_{tt}(\mathbf{x},t)=\Delta u(\mathbf{x},t)+\delta
(z-z_{0})f(t),\ (\mathbf{x},t)\in \mathbb{R}^{3}\times (0,\infty ),
\label{eq:fp1} \\
&&u(\mathbf{x},0)=0,\quad u_{t}(\mathbf{x},0)=0,  \label{eq:fp2}
\end{eqnarray}%
where $u:=E_{2}$ is the wave generated by the incident plane wave
propagating along the $z$-axis and incident at the plane $\left\{
z=z_{0}\right\} $. Here $f\left( t\right) \not\equiv 0$ is a piecewise
continuous bounded function representing the time-dependent waveform of the
incident plane wave. It was demonstrated numerically in \cite%
{Beilina:CEJM2013} that the component $E_{2}$ dominates components $%
E_{1},E_{3}$ and that the propagation of $E_{2}$ is governed well by the
solution of the Cauchy problem (\ref{eq:fp1})--(\ref{eq:fp2}).

The coefficient $\epsilon (\mathbf{x})$ in (\ref{eq:fp1}) represents the
spatially distributed dielectric constant of the medium in which the wave
propagates. Let $\Omega \subset \mathbb{R}^{3}$ be a bounded domain and let
the plane where the incident wave is emitted be located outside of the
domain $\overline{\Omega }$, i.e., $\overline{\Omega }\cap
\{z=z_{0}\}=\varnothing $. We assume that there exist two positive constants 
$0<\epsilon _{l}\leq 1$ and $\epsilon _{u}\geq 1$ such that 
\begin{equation}
\epsilon _{l}\leq \epsilon (\mathbf{x})\leq \epsilon _{u},\ \forall \mathbf{x%
}\in \mathbb{R}^{3},\quad \epsilon (\mathbf{x})\equiv 1,\ \forall \mathbf{x}%
\notin \Omega .  \label{eq:fp3}
\end{equation}%
In other words, the medium is assumed to be homogeneous outside of $\Omega $%
. We note that this assumption is not true in the case of objects buried in
the ground since the dielectric constant of the ground is different from
that of the air. Therefore, to use our model in this case, we first
preprocess the experimental data, and then treat the buried objects as they
are in air.

Below we denote by $C^{k+\alpha }$ the H\"{o}lder spaces, where $k\geq 0$ is
an integer and $\alpha \in \left( 0,1\right) $. Let $d>2\epsilon _{u}$ be a
constant. In addition to (\ref{eq:fp3}), in our theoretical analysis we
assume that the function $\epsilon (\mathbf{x})$ is unknown inside $\Omega $
and 
\begin{equation}
\epsilon \in C^{\alpha }\left( \mathbb{R}^{3}\right) , \quad \| \epsilon
\|_{C^{\alpha }\left( \mathbb{R}^{3}\right) }<d,  \label{2.0}
\end{equation}
where $d >1$ is a given constant. In this work, we consider the following
CIP:

\textbf{CIP}: \textit{Reconstruct the coefficient $\epsilon (\mathbf{x})$\
for $\mathbf{x}\in \Omega $, given the following measured data for a single
incident plane wave generated at the plane $\{z=z_{0}\}$ outside of $%
\overline{\Omega }$,} 
\begin{equation}
g(\mathbf{x},t)=u(\mathbf{x},t),\ \mathbf{x}\in \partial \Omega ,\ t\in
(0,\infty ).  \label{eq:mea}
\end{equation}

For the theoretical analysis, we state the inverse problem for the case when
the data are given at the entire boundary and the boundary $\partial \Omega $
being $C^{3}$-regular. However, only the backscattering data are measured in
our experiment and we use a rectangular prism as $\Omega $ in our
computations. As in our previous works, we complete the missing data by the
solution of the forward model (\ref{eq:fp1})--(\ref{eq:fp2}) in a
homogeneous medium, see section \ref{sec:num}.

The assumption of the infinite time interval in (\ref{eq:mea}) is not
restrictive, because in our method we apply the Laplace transform to $g(%
\mathbf{x,}t)$ with respect to the time variable $t$. Since the kernel of
this transform decays exponentially with respect to $t$, the Laplace
transform effectively cuts off to zero values of the function $g(\mathbf{x,}%
t)$ for large $t$. Moreover, since the incident wave is excited for a finite
time interval, our experimental observation has always been that the total
wave almost vanishes after a finite time interval, too.

Concerning the uniqueness of this CIP, global uniqueness theorems for
multidimensional CIPs with a single measurement are currently known only
under the assumption that at least one initial condition does not equal zero
in the entire domain $\overline{\Omega }$. Proofs of such theorems are based
on Carleman estimates, see \cite{BukhKlib,KT:2004} and sections 1.10, 1.11
in \cite{B-K:2012}. Since both initial conditions (\ref{eq:fp2}) equal zero
in $\overline{\Omega },$ this method is inapplicable to our case. However,
since we need to solve numerically our CIP anyway, we assume that the
uniqueness holds.

We remark that (\ref{eq:fp1}) is invalid if metallic objects are present in
the domain $\Omega $. To deal with this type of targets, we follow a
suggestion of \cite{K-B-K-S-N-F:IP2012}. It was established numerically in 
\cite{K-B-K-S-N-F:IP2012} that metals can be modeled as dielectrics with a
high dielectric constant, which is referred to as the \textit{effective}
dielectric constant of metals.

\subsection{The globally convergent method in brief}

\label{sec:gca}

The globally convergent method of \cite{B-K:2012} works with the Laplace
transformed data. However, we do not invert the Laplace transform. Let 
\begin{equation}
\tilde{u}(\mathbf{x},s):=\left( \mathcal{L}u\right) (\mathbf{x}%
,s)=\int\limits_{0}^{\infty }u\left( \mathbf{x},t\right) e^{-st}dt,\ s\geq 
\underline{s}=\underline{s}\left( d\right) >0,  \label{200}
\end{equation}%
and $\tilde{f}(s)=\left( \mathcal{L}f\right) (s)$ be the Laplace transform
of $u(x,t)$ and $f(t)$, respectively, where $s$ is referred to as the pseudo
frequency. We assume that $s\geq \underline{s}(d)>0$, where the number $%
\underline{s}\left( d\right) $ is large enough so that the Laplace
transforms of the function $u$ and its derivatives $D^{\beta }u,\left\vert
\beta \right\vert =1,2,$ converge absolutely. The number $d$ is defined in (%
\ref{2.0}). We assume that $\tilde{f}(s)\neq 0$ for all $s\geq \underline{s}%
(d)$. Define $w(\mathbf{x},s):=\tilde{u}(\mathbf{x},s)/\tilde{f}(s)$. Then,
this function satisfies the equation: 
\begin{equation}
\Delta w(\mathbf{x},s)-s^{2}\epsilon (\mathbf{x})w(\mathbf{x},s)=-\delta
(z-z_{0}),\ \mathbf{x}\in \mathbb{R}^{3},\ s\geq \underline{s}\left(
d\right) .  \label{eq:w}
\end{equation}%
Define 
\begin{equation}
w_{0}\left( z,s\right) =\frac{\exp \left( -s\left\vert z-z_{0}\right\vert
\right) }{2s}.  \label{201}
\end{equation}%
The function $w_{0}\left( z,s\right) $ is the unique solution equation (\ref%
{eq:fp1}) for the case $\epsilon \left( \mathbf{x}\right) \equiv 1,$ which
tends to zero as $\left\vert z\right\vert \rightarrow \infty .$ It is shown
in Theorem \ref{th:1} (section \ref{sec:3}) that in the case $f\left(
t\right) =\delta \left( t\right) $   
\begin{equation}
\lim_{\left\vert \mathbf{x}\right\vert \rightarrow \infty }\left[ w\left( 
\mathbf{x},s\right) -w_{0}(z,s)\right] =0  \label{1}
\end{equation}%
and that the function $w\left( \mathbf{x},s\right) $ can be represented in
the form%
\begin{equation}
w\left( \mathbf{x},s\right) =w_{0}\left( z,s\right) +\widehat{w}\left( 
\mathbf{x},s\right) ,\text{ where }\widehat{w}\left( \mathbf{x},s\right) \in
C^{2+\alpha }\left( \mathbb{R}^{3}\right) ,\ \forall s\geq \underline{s}%
\left( d\right) .  \label{2}
\end{equation}%
Furthermore, the same theorem claims that $w(\mathbf{x},s)>0.$ Thus, we
assume these properties in our algorithm even if $f\left( t\right) \neq
\delta \left( t\right) .$ Next, define the function $v$ by $v:=(\ln w)/s^{2}$%
. 
Substituting $w=e^{vs^{2}}$ into (\ref{eq:w}) and keeping in mind that $%
\overline{\Omega }\cap \{z=z_{0}\}=\varnothing $, we obtain 
\begin{equation}
\Delta v+s^{2}|\nabla v|^{2}=\epsilon (\mathbf{x}),\mathbf{x}\in \Omega .
\label{eq:c}
\end{equation}%
Equation (\ref{eq:c}) shows that the coefficient $\epsilon (\mathbf{x})$ can
be computed directly via the function $v$. We now eliminate the
unknown coefficient $\epsilon (x)$ from equation (\ref{eq:c}) via
the differentiation with respect to $s$, which is similar to the
first step of the method of \cite{BukhKlib,KT:2004}. Define the function $q$
by $q:=\partial v/\partial s$. Then $v=-\int\limits_{s}^{\infty }qd\tau $.
It follows from (\ref{eq:c}) that $q$ satisfies the following integral
differential equation: 
\begin{equation}
\Delta q-2s^{2}\nabla q\cdot \int\limits_{s}^{\infty }\nabla q(\mathbf{x}%
,\tau )d\tau +2s\left\vert \int\limits_{s}^{\infty }\nabla q(\mathbf{x},\tau
)d\tau \right\vert ^{2}=0,\ \mathbf{x}\in \Omega .  \label{eq:q0}
\end{equation}%
Moreover, it follows from (\ref{eq:mea}) that $q$ satisfies the following
boundary condition: 
\begin{equation}
q(\mathbf{x},s)=\psi (\mathbf{x},s),\ \mathbf{x}\in \partial \Omega ,
\label{eq:q2}
\end{equation}%
where $\psi $ is derived from the boundary measured data by $\psi (\mathbf{x}%
,s)=\frac{\partial }{\partial s}\left[ \frac{\ln (\varphi )}{s^{2}}\right] $
with $\varphi (\mathbf{x},s)=\int\limits_{0}^{\infty }g(\mathbf{x}%
,t)e^{-st}dt/\tilde{f}(s).$ To solve the problem (\ref{eq:q0}), (\ref{eq:q2}%
) for $q$, we use the following approach. We represent the integral over the
infinite interval as 
\begin{equation}
v=-\int\limits_{s}^{\infty }qd\tau =-\int\limits_{s}^{\bar{s}}qd\tau +V,
\label{eq:v2}
\end{equation}%
where $\bar{s}>\underline{s}$, which plays the role of a regularization
parameter and is chosen numerically in the computational practice. The
function $V(\mathbf{x}):=v(\mathbf{x},\bar{s})$ is called the
\textquotedblleft tail function.\textquotedblright\ Note that 
\begin{equation}
V(\mathbf{x})=\frac{\ln w(\mathbf{x},\bar{s})}{\bar{s}^{2}}.  \label{eq:tail}
\end{equation}%
From (\ref{eq:q0}) and (\ref{eq:v2}) we obtain the following nonlinear
integral differential equation involving $q$ and $V$: 
\begin{eqnarray}
\Delta q &-&2s^{2}\nabla q\cdot \int\limits_{s}^{\bar{s}}\nabla q(\mathbf{x}%
,\tau )d\tau +2s^{2}\nabla V\cdot \nabla q+2s\left\vert \int\limits_{s}^{%
\bar{s}}\nabla q(\mathbf{x},\tau )d\tau \right\vert ^{2}  \notag \\
&-&4s\nabla V\cdot \int\limits_{s}^{\bar{s}}\nabla q(\mathbf{x},\tau )d\tau
+2s\left\vert \nabla V\right\vert ^{2}=0,\ \mathbf{x}\in \Omega .
\label{eq:q}
\end{eqnarray}%
Note that (\ref{eq:q}) has two unknown functions $q$ and $V$. In order to
approximate both of them, we use a predictor-corrector-type approach:
starting from an initial guess of the tail function $V$, we solve the
problem (\ref{eq:q}), (\ref{eq:q2}) for $q$. After that, we calculate the
coefficient $\epsilon $ via (\ref{eq:c}) and solve the forward problem for $u
$. Next, we update the tail function $V$ via (\ref{eq:tail}) and repeat the
iterative procedure again. Thus, $V$ is the `predictor' and $q$ is the
`corrector' here.

To approximate the integrals in (\ref{eq:q}), we make use of a layer
stripping procedure with respect to $s$ described as follows. Divide the
pseudo frequency interval $[\underline{s},\bar{s}]$ into $N$ uniform
subintervals by $\bar{s}=s_{0}>s_{1}>\cdots >s_{N}=\underline{s},\
s_{n}-s_{n+1}=h.$ We approximate $q$ by a piece-wise constant function: $q(%
\mathbf{x},s)\approx q_{n}(\mathbf{x}),\ s\in (s_{n},s_{n-1}],\ n=1,\dots ,N.
$ We also set $q_{0}\equiv 0$. Then after some manipulations, a system of
elliptic equations for functions $q_{n}\left( x\right) $ is derived from (%
\ref{eq:q}) using the so-called \textquotedblleft Carleman Weight Function" $%
\exp \left[ \lambda \left( s-s_{n-1}\right) \right] ,\ s\in \left(
s_{n},s_{n-1}\right) ,$ where $\lambda \gg 1$ is a certain parameter. This
system is \cite{TBKF:SISC2014} 
\begin{eqnarray}
\Delta q_{n} &+&A_{1,n}\nabla q_{n}\cdot \left( \nabla V_{n}-\nabla 
\overline{q_{n-1}}\right)   \notag \\
&=&A_{2,n}|\nabla q_{n}|^{2}+A_{3,n}\left( |\nabla \overline{q_{n-1}}%
|^{2}+|\nabla V_{n}|^{2}-2\nabla V_{n}\cdot \nabla \overline{q_{n-1}}\right)
,  \label{eq:q5}
\end{eqnarray}%
where $A_{i,n},\ i=1,2,3,$ are some coefficients, depending on $s_{n}$ and $%
\lambda $, which can be analytically computed, and $\nabla \overline{q_{n-1}}%
=h\sum_{j=0}^{n-1}\nabla q_{j}$. Here we indicate the dependence of the tail
function $V:=V_{n}$ on the number $n$, because we approximate $V$
iteratively. The discretized version of the boundary condition (\ref{eq:q2})
is given by 
\begin{equation}
q_{n}(\mathbf{x})=\psi _{n}(\mathbf{x}):=\frac{1}{h}\int%
\limits_{s_{n}}^{s_{n-1}}\psi (\mathbf{x},s)ds\approx \frac{1}{2}[\psi (%
\mathbf{x},s_{n})+\psi (\mathbf{x},s_{n-1})],\ \mathbf{x}\in \partial \Omega
.  \label{eq:q7}
\end{equation}%
One can prove that $\left\vert A_{2,n}\right\vert \leq C/\lambda $ for
sufficiently large $\lambda ,$ where $C>0$ is a certain constant. Hence, the
first term in the right-hand side of (\ref{eq:q5}) is dominated by the other
terms. Therefore, in the following we set $A_{2,n}|\nabla q_{n}|^{2}:=0$.
The system of elliptic equations (\ref{eq:q5}) with boundary conditions (\ref%
{eq:q7}) can be solved sequentially starting from $n=1$. To solve it, we
make use of the iterative process: For a given $n$ and some approximation $%
q_{n,i-1}$ of $q_{n}$, we find the next approximation $q_{n,i}$ of $q_{n}$
by solving the Dirichlet boundary value problem (\ref{eq:q5})--(\ref{eq:q7})
with $q_{n}$ on the right-hand side replaced by $q_{n,i-1}$. Denote by $m_{n}
$ the number of these iterations.

Let $\Omega ^{\prime }\subset \Omega $ be a certain subdomain with $\partial
\Omega ^{\prime }\cap \partial \Omega =\varnothing .$ This subdomain is
chosen computationally.\textbf{\ }Choose a function $\chi \left( \mathbf{x}%
\right) \in C^{1}\left( \mathbb{R}^{3}\right) $ such that%
\begin{equation*}
\chi \left( \mathbf{x}\right) =\left\{ 
\begin{array}{c}
1,\mathbf{x}\in \Omega ^{\prime }, \\ 
\in \left[ 0,1\right] ,\mathbf{x}\in \Omega \diagdown \Omega ^{\prime }, \\ 
0,\mathbf{x}\in \mathbb{R}^{3}\diagdown \Omega .%
\end{array}%
\right. 
\end{equation*}%
The steps of the globally convergent algorithm is summarized as follows.

\begin{itemize}
\item Given the first tail $V_{1,1}:=V_{0}$. Set $q_{0}\equiv 0$.

\item For $n = 1, 2, \dots, N$

\begin{enumerate}
\item Set $q_{n,0}=q_{n-1}$, $V_{n,1}=V_{n-1}.$

\item For $i = 1, 2,\dots, m_n$

\begin{itemize}
\item Find $q_{n,i}$ by solving the problem (\ref{eq:q5}), (\ref{eq:q7})
with $V_{n}:=V_{n,i}$.

\item Compute $v_{n,i} = -h q_{n,i} - \overline{q_{n-1}} + V_{n,i},\ \mathbf{%
x} \in \Omega$.

\item Compute $\epsilon _{n,i}$ via (\ref{eq:c}). Then set $\overline{%
\epsilon }_{n,i}\left( \mathbf{x}\right) =\left( 1-\chi \left( \mathbf{x}%
\right) \right) +\chi \left( \mathbf{x}\right) \epsilon _{n,i}\left( \mathbf{%
x}\right) .$ Next, solve the forward problem (\ref{eq:fp1}), (\ref{eq:fp2})
with the new computed coefficient $\epsilon :=\overline{\epsilon }_{n,i}$,
compute $w:=w_{n,i}$ and update the tail $V_{n,i+1}$ via (\ref{eq:tail}).
\end{itemize}

\item Set $q_{n}=q_{n,m_{n}}$, $\epsilon _{n}=\epsilon _{n,m_{n}},\
V_{n}=V_{n,m_{n}+1}$ and go to the next frequency interval $\left[
s_{n+1},s_{n}\right] $ if $n<N.$ If one of the stopping criteria is
satisfied at $n:=\overline{N}\in \left[ 1,N\right] $ or $n=N$, then stop.
\end{enumerate}
\end{itemize}

In computations, the number of inner iterations $m_{n}$ is determined by the
stopping criterion with respect to $i$. In this paper, we use the stopping
criteria with respect to $i,n$ proposed in \cite{BTKF:IP2013, TBKF:SISC2014}%
, see a brief description in Section \ref{sec:num}.

\subsection{The initial tail function}

\label{sec:gc}

We remark that the convergence of this algorithm depends on the choice of
the initial tail function $V_{1,1}$. To ensure the global convergence, the
choice of this function should not rely on any a priori knowledge of the
exact solution of our CIP. First, in accordance with the Tikhonov approach
to ill-posed problems (see, e.g., section 1.4 in \cite{B-K:2012}), we assume
the existence of the exact solution $\epsilon ^{\ast }\left( \mathbf{x}%
\right) $ of our CIP for noiseless data $g^{\ast }\left( \mathbf{x},t\right) 
$ in (\ref{eq:mea}). We assume that the function $\epsilon ^{\ast }\left( 
\mathbf{x}\right) $ satisfies conditions (\ref{eq:fp3}), (\ref{2.0}). Let $%
V^{\ast }(\mathbf{x},s)$ and $w^{\ast }(\mathbf{x},s)$ be respectively the
tail function and the function $w$ which correspond to $\epsilon ^{\ast
}\left( \mathbf{x}\right) $. In \cite{Romanov:MJM2006} the solution of a
general hyperbolic equation is constructed for the case when the wave is
generated by the plane wave. Using results of \cite{Romanov:MJM2006}, one
can prove, that under some conditions, there exists a function $p^{\ast }(%
\mathbf{x})\in C^{2+\alpha }\left( \overline{\Omega }\right) $ such that $%
V^{\ast }(\mathbf{x},s)=p^{\ast }(\mathbf{x})/s+O\left( 1/s^{2}\right)
,s\rightarrow \infty $. The proof is similar to the proof of an analogous
formula (5.3.17) in \cite{KT:2004}, which has used the construction of \cite%
{Romanov:1986} of the solution of a general hyperbolic equation for the case
of the point source.

Due to this asymptotic behavior, we introduce our approximate mathematical
model via the assumption that the exact tail is given by 
\begin{equation}
V^{\ast }(\mathbf{x},s)=\frac{p^{\ast }\left( \mathbf{x}\right) }{s}=\frac{%
\ln w^{\ast }\left( \mathbf{x},s\right) }{s^{2}},\text{ }\forall s\geq 
\overline{s}.  \label{eq:model2}
\end{equation}%
Since $q\left( \mathbf{x},s\right) =\partial _{s}V\left( \mathbf{x},s\right)
,$ then (\ref{eq:model2}) implies that $q^{\ast }\left( \mathbf{x},s\right)
=-p^{\ast }(\mathbf{x})/s^{2},$ $\forall s\geq \overline{s}.$ Note that we
use this assumption only on the initializing iteration to obtain $%
V_{1,1}\left( \mathbf{x}\right) .$ Substituting these expressions for $%
V^{\ast },\ q^{\ast }$ at $s=\overline{s}$ in (\ref{eq:q2}), (\ref{eq:q}),
we obtain that the function $p^{\ast }\left( \mathbf{x}\right) $ satisfies 
\begin{eqnarray}
\Delta p^{\ast }(\mathbf{x}) &=&0,\ \mathbf{x}\in \Omega ,\quad p^{\ast }\in
C^{2+\alpha }\left( \overline{\Omega }\right) ,  \label{3.37} \\
p^{\ast }|_{\partial \Omega } &=&-\overline{s}^{2}\psi ^{\ast }\left( 
\mathbf{x},\overline{s}\right) ,  \label{3.38}
\end{eqnarray}%
where $\psi ^{\ast }\left( \mathbf{x},s\right) $ is the function in (\ref%
{eq:q2}) for the case of the exact data. Since actually we have the function 
$\psi \left( \mathbf{x},s\right) ,$ which is supposed to be contaminated by
noise, then as the first guess for the tail function we take 
\begin{equation}
V_{1,1}\left( \mathbf{x}\right) :=\frac{p\left( \mathbf{x}\right) }{%
\overline{s}},\ \mathbf{x}\in \Omega ,  \label{3.43}
\end{equation}%
where the function $p\left( \mathbf{x}\right) $ is the solution of the
problem (\ref{3.37}), (\ref{3.38}) for the case when $\psi ^{\ast }\left( 
\mathbf{x},\overline{s}\right) $ is replaced by $\psi \left( \mathbf{x},%
\overline{s}\right) ,$%
\begin{eqnarray}
\Delta p(\mathbf{x}) &=&0,\ \mathbf{x}\in \Omega ,  \label{3.44} \\
p|_{\partial \Omega } &=&-\overline{s}^{2}\psi \left( \mathbf{x},\overline{s}%
\right) .  \label{3.45}
\end{eqnarray}

\section{Global convergence}

\label{sec:3}

In this section we briefly discuss the question of the approximate global
convergence of the above numerical method. The only difference between this
convergence analysis and the one in Theorem 2.9.4 in \cite{B-K:2012} and
Theorem 5.1 in \cite{B-K:JIIP2012} is that we now use the plane wave instead
of the point source. This causes the difference only in the proof of Theorem
2.7.2 of \cite{B-K:2012}. Theorem \ref{th:1} is a direct analog of Theorem
2.7.2 of \cite{B-K:2012} for the case of the plane wave. For brevity we
point here only to those features of the proof of Theorem \ref{th:1} which
are different from those of Theorem 2.7.2 of \cite{B-K:2012}. Next, for the
convenience of the reader, we briefly outline the proof of Theorem \ref{th:2}%
, since it is completely the same as proofs of Theorem 5.1 in \cite%
{B-K:JIIP2012} and Theorem 2.9.4 in \cite{B-K:2012}. To follow more closely
arguments of \cite{B-K:2012, B-K:JIIP2012}, we assume in our convergence
analysis that the lower bound $\epsilon _{l}$ in (\ref{eq:fp3}) is chosen by 
\begin{equation}
\epsilon _{l}=1.  \label{4.1}
\end{equation}

\begin{theorem}\label{th:1}
Let the function $\epsilon \left( \mathbf{x}%
\right) $ satisfies conditions (\ref{eq:fp3}), (\ref{2.0}) and (\ref{4.1})
and let in (\ref{eq:fp1}) $f\left( t\right) =\delta \left( t\right) $. Then there exists a number $\underline{s}\left( d\right) >0$ such
that for every $s>\underline{s}\left( d\right) $ the function $w\left( 
\mathbf{x},s\right) =\mathcal{L}\left( u\right) \left( \mathbf{x},s\right) $ 
satisfies conditions (\ref{eq:w})--(\ref{2}), where the operator $\mathcal L$
 is defined in (\ref{200}). Furthermore, 
\begin{equation}
w_{u}\left( z,s\right) <w\left( \mathbf{x},s\right) \leq w_{0}\left(
z,s\right) ,  \label{4.3}
\end{equation}%
where the function $w_{0}\left( z,s\right) $ is defined in (\ref{201}) and 
$w_{u}\left( z,s\right) $ is such a solution of (\ref{eq:w}) for $%
\epsilon \left( \mathbf{x}\right) \equiv \epsilon _{u}$ which tends to zero
as $\left\vert z\right\vert \rightarrow \infty ,$ i.e., 
\begin{equation*}
w_{u}\left( z,s\right) =\exp \left( -s\sqrt{\epsilon _{u}}\left\vert
z-z_{0}\right\vert \right) /\left( 2s\sqrt{\epsilon _{u}}\right) .
\end{equation*}
Also, the solution of the problem (\ref{eq:w})--(\ref{2}) is unique.
\end{theorem}


%
%

\textit{Proof.} Note that the classical theory of elliptic equations \cite%
{GT:1983} is not directly applicable here, since it works only with bounded
domains, whereas we work here with (\ref{eq:w}) in the entire space $\mathbb{%
R}^{3}.$ This causes additional difficulties of the proof. In this proof, we
consider $s>\underline{s}\left( d\right) .$

It is well known that in the case \textit{\ }$f\left( t\right) =\delta
\left( t\right) $ the Cauchy problem (\ref{eq:fp1}), (\ref{eq:fp2}) is
equivalent to the Cauchy problem for the homogeneous equation (\ref{eq:fp1})
with initial conditions $u\left( x,0\right) =0,u_{t}\left( x,0\right)
=\delta \left( z-z_{0}\right) $ \cite{V}. Consider the well known
Laplace-like transform, which transforms the Cauchy problem for the
hyperbolic equation in the Cauchy problem for a similar parabolic equation
(see, e.g., formula (1.162) in \cite{B-K:2012}), 
\begin{equation*}
v\left( \mathbf{x},t\right) :=\left( \mathcal{L}_{1}u\right) \left( \mathbf{x%
},t\right) :=\frac{1}{2\sqrt{\pi }t^{3/2}}\int\limits_{0}^{\infty }u\left( 
\mathbf{x},\tau \right) \tau \exp \left( -\frac{\tau ^{2}}{4t}\right) d\tau
,\ t>0.
\end{equation*}%
Then $v\left( \mathbf{x},t\right) $ is the solution of the following
parabolic Cauchy problem%
\begin{eqnarray}
\epsilon \left( \mathbf{x}\right) v_{t} &=&\Delta v,\text{ }\left( \mathbf{x}%
,t\right) \in \mathbb{R}^{3}\times \left( 0,\infty \right) ,  \label{4.4} \\
v\left( \mathbf{x},0\right)  &=&\delta \left( z-z_{0}\right) .  \label{4.5}
\end{eqnarray}%
Also, define another analog of the Laplace transform,%
\begin{equation}
\left( \mathcal{L}_{2}v\right) \left( \mathbf{x},s\right)
=\int\limits_{0}^{\infty }v\left( \mathbf{x},t\right) e^{-s^{2}t}dt,s>0.
\label{202}
\end{equation}%
By the formula (28) of section 4.5 of tables of the Laplace transform \cite%
{BE:1954} 
\begin{equation*}
\int\limits_{0}^{\infty }e^{-s^{2}t}\left[ \frac{1}{2\sqrt{\pi }t^{3/2}}\tau
\exp \left( -\frac{\tau ^{2}}{4t}\right) \right] dt=e^{-s\tau };\ s,\tau >0.
\end{equation*}%
Hence, using (\ref{200}) and (\ref{202}), we obtain for sufficiently large $%
s>0$ 
\begin{equation}
w\left( \mathbf{x},s\right) =\left( \mathcal{L}u\right) \left( \mathbf{x}%
,s\right) =\mathcal{L}_{2}\left( \mathcal{L}_{1}u\right) \left( \mathbf{x}%
,s\right) =\left( \mathcal{L}_{2}v\right) \left( \mathbf{x},s\right) .
\label{203}
\end{equation}%
Hence, we study now the problem (\ref{4.4}), (\ref{4.5}). Let $Z\left( 
\mathbf{x,\xi },t,\tau \right) $ be the fundamental solution for the
parabolic operator $\epsilon \left( \mathbf{x}\right) \partial _{t}-\Delta .$
Then the formula (13.1) of Chapter 4 of \cite{L-S-U:1968} along with other
detailed estimates  of the fundamental solution of the general second order
parabolic equation given in \S 11-\S 14 of Chapter 4 of \cite{L-S-U:1968}
imply that for $2r+\left\vert \gamma \right\vert \leq 2,t>\tau $ 
\begin{equation}
\left\vert D_{t}^{r}D_{\mathbf{x}}^{\gamma }Z\left( \mathbf{x,\xi },t,\tau
\right) \right\vert \leq \frac{\exp \left( Mt\right) }{\left( t-\tau \right)
^{\left( 3+2r+\left\vert \gamma \right\vert \right) /2}}\exp \left( -C\frac{%
\left\vert \mathbf{x-\xi }\right\vert ^{2}}{t-\tau }\right) .  \label{4.6}
\end{equation}%
Here and below$M=M\left( d\right) >0$ and $C=C\left( d\right) >0$ denote
different numbers depending only on the number $d$. By (\ref{4.4}), (\ref%
{4.5}) 
\begin{equation*}
v\left( \mathbf{x},t\right) =\int\limits_{\mathbb{R}^{3}}Z\left( \mathbf{%
x,\xi },t,0\right) \delta \left( \xi _{3}-z_{0}\right) d\mathbf{\xi =}%
\int\limits_{\mathbb{R}^{2}}Z\left( \mathbf{x,}\xi _{1},\xi
_{2},z_{0},t,0\right) d\xi _{1}d\xi _{2}.
\end{equation*}%
Hence, (\ref{4.6}) implies that 
\begin{equation*}
\left\vert D_{t}^{r}D_{\mathbf{x}}^{\gamma }v\left( \mathbf{x},t\right)
\right\vert \leq \frac{\exp \left( Mt\right) }{t^{\left( 1+2r+\left\vert
\gamma \right\vert \right) /2}}\exp \left( -C\frac{\left( z-z_{0}\right) ^{2}%
}{t}\right) .
\end{equation*}%
Hence, one can apply the operator $\mathcal{L}_{2}$ to the functions $%
D_{t}^{r}D_{\mathbf{x}}^{\gamma }v\left( \mathbf{x},t\right) $ for $s\geq 
\underline{s}\left( d\right) =\sqrt{M\left( d\right) }.$ Hence, (\ref{4.4}),
(\ref{4.5}) and (\ref{203}) imply that the function $w$ satisfies (\ref{eq:w}%
).

We now prove that the function $w$ satisfies conditions (\ref{1}), (\ref{2}%
). Let 
$$v_{0}\left( z,t\right) =\exp \left[ -\left( z-z_{0}\right)
^{2}/\left( 4t\right) \right] /\left( 2\sqrt{\pi t}\right) $$ be the solution
of the problem (\ref{4.4}), (\ref{4.5}) for the case $\epsilon \left( 
\mathbf{x}\right) \equiv 1.$ Denote $\widetilde{v}\left( \mathbf{x},t\right)
=v\left( \mathbf{x},t\right) -v_{0}\left( z,t\right) .$ Then 
\begin{equation}
\epsilon \left( \mathbf{x}\right) \widetilde{v}_{t}-\Delta \widetilde{v}=%
\left[ 1-\epsilon \left( \mathbf{x}\right) \right] v_{0t},\quad \widetilde{v}%
\left( \mathbf{x},0\right) =0.  \label{4.8}
\end{equation}%
Since $\epsilon \left( \mathbf{x}\right) =1$ for $\mathbf{x}\notin \Omega $
and $\Omega \cap \left\{ z=z_{0}\right\} =\varnothing ,$ then the right-hand
side of equation (\ref{4.8}) does not have a singularity. Hence, (\ref{4.8})
implies that%
\begin{equation*}
\widetilde{v}\left( \mathbf{x},t\right) =\int\limits_{0}^{t}d\tau
\int\limits_{\Omega }Z\left( \mathbf{x,\xi },t,\tau \right) \left[
1-\epsilon \left( \mathbf{\xi }\right) \right] v_{0\tau }\left( \xi
_{3},\tau \right) d\mathbf{\xi }.
\end{equation*}%
Hence, it follows from (\ref{4.6}) that 
\begin{equation}
\left\vert \widetilde{v}\left( \mathbf{x},t\right) \right\vert \leq \exp
\left( Mt\right) \int\limits_{0}^{t}d\tau \int\limits_{\Omega }\frac{1}{%
\left( t-\tau \right) ^{3/2}}\exp \left( -C\frac{\left\vert \mathbf{x-\xi }%
\right\vert ^{2}}{t-\tau }\right) \left\vert v_{0\tau }\left( \xi _{3},\tau
\right) \right\vert d\mathbf{\xi }.  \label{4.10}
\end{equation}%
Also, 
\begin{equation}
\left\vert v_{0\tau }\left( \xi _{3},\tau \right) \right\vert \leq C_{1}
\left[ \frac{\left( \xi _{3}-z_{0}\right) ^{2}}{\tau ^{5/2}}+\frac{1}{\tau
^{3/2}}\right] \exp \left[ -\frac{\left( \xi _{3}-z_{0}\right) ^{2}}{4\tau }%
\right] ,  \label{4.11}
\end{equation}%
where the number $C_{1}>0$ is independent of $\xi _{3},z_{0},\tau .$ Using
formulas (28) and (29) of section 4.5 of \cite{BE:1954}, we obtain 
\begin{equation}
\mathcal{L}_{2}\left[ \frac{1}{t^{3/2}}\exp \left( -C\frac{\left\vert 
\mathbf{x-\xi }\right\vert ^{2}}{t}\right) \right] =\frac{\sqrt{\pi }\exp %
\left[ -2\sqrt{C}s\left\vert \mathbf{x-\xi }\right\vert \right] }{\sqrt{C}%
\left\vert \mathbf{x-\xi }\right\vert },  \label{4.12}
\end{equation}%
\begin{equation}
\mathcal{L}_{2}\left\{ \frac{1}{t^{3/2}}\exp \left[ -\frac{\left( \xi
_{3}-z_{0}\right) ^{2}}{4t}\right] \right\} =\frac{2\sqrt{\pi }}{\left\vert
\xi _{3}-z_{0}\right\vert }\exp \left( -s\left\vert \xi
_{3}-z_{0}\right\vert \right) ,  \label{4.13}
\end{equation}%
\begin{equation}
\mathcal{L}_{2}\left\{ \frac{\left( \xi _{3}-z_{0}\right) ^{2}}{t^{5/2}}\exp %
\left[ -\frac{\left( \xi _{3}-z_{0}\right) ^{2}}{4t}\right] \right\} =\frac{%
16s^{3}}{\left\vert \xi _{3}-z_{0}\right\vert }K_{3/2}\left( s\left\vert \xi
_{3}-z_{0}\right\vert \right) ,  \label{4.14}
\end{equation}%
where $K_{3/2}$ is the McDonald function. Since $\overline{\Omega }\cap
\left\{ z=z_{0}\right\} =\varnothing ,$ then the right-hand sides of
formulas (\ref{4.13}) and (\ref{4.14}) do not have a singularity for $%
\mathbf{\xi }\in \Omega .$

By the formula (27) of section 4.5 of \cite{BE:1954} $\left( \mathcal{L}%
_{2}v_{0}\right) \left( z,s\right) =w_{0}\left( z,s\right) .$ Hence, let $%
\widehat{w}\left( \mathbf{x},s\right) =w\left( \mathbf{x},s\right)
-w_{0}\left( z,s\right) =\left( \mathcal{L}_{2}\widetilde{v}\right) \left( 
\mathbf{x},s\right) .$ Since $\left\vert \left( \mathcal{L}_{2}\widetilde{v}%
\right) \left( \mathbf{x},s\right) \right\vert \leq \left( \mathcal{L}%
_{2}\left( \left\vert \widetilde{v}\right\vert \right) \right) \left( 
\mathbf{x},s\right) ,$ then (\ref{4.10})--(\ref{4.14}) and the convolution
theorem for the Laplace transform imply that 
\begin{equation*}
\begin{split}
\left\vert \widehat{w}\left( \mathbf{x},s\right) \right\vert & \leq
C_{2}s^{3}\int\limits_{\Omega }\frac{\exp \left( -2\sqrt{C}s\left\vert 
\mathbf{x}-\mathbf{\xi }\right\vert \right) }{\left\vert \mathbf{x}-\mathbf{%
\xi }\right\vert }\left[ \frac{\exp \left( -s\left\vert \xi
_{3}-z_{0}\right\vert \right) }{\left\vert \xi _{3}-z_{0}\right\vert }+\frac{%
K_{3/2}\left( s\left\vert \xi _{3}-z_{0}\right\vert \right) }{\left\vert \xi
_{3}-z_{0}\right\vert }\right] d\mathbf{\xi } \\
& \leq C_{2}\exp \left( -C_{3}\left( s-\sqrt{M\left( d\right) }\right)
\left\vert \mathbf{x}\right\vert \right) ,\ s>\underline{s}\left( d\right) =%
\sqrt{M\left( d\right) },\ \left\vert \mathbf{x}\right\vert \rightarrow
\infty ,
\end{split}%
\end{equation*}%
where numbers $C_{2}=C_{2}\left( C,\Omega ,z_{0}\right) >0,\
C_{3}=C_{3}\left( C,\Omega ,z_{0}\right) >0$ depend only on the listed
parameters. Therefore, condition (\ref{1}) holds.

Next, the function $\widehat{w}\left( \mathbf{x},s\right) $ satisfies the
following conditions: 
\begin{eqnarray}
\Delta \widehat{w}-s^{2}\epsilon \left( \mathbf{x}\right) \widehat{w}
&=&s^{2}\left( \epsilon \left( \mathbf{x}\right) -1\right) w_{0}\left(
z,s\right) ,\text{ }s>\underline{s}\left( d\right) =\sqrt{M\left( d\right) },
\label{4.15} \\
\lim_{\left\vert x\right\vert \rightarrow \infty }\widehat{w}\left( \mathbf{x%
},s\right) &=&0.  \label{4.16}
\end{eqnarray}%
Since the right-hand side of (\ref{4.15}) belongs to $C^{\alpha }\left( 
\mathbb{R}^{3}\right) ,$ then $\widehat{w}\left( \mathbf{x},s\right) \in
C^{2+\alpha }\left( \overline{G}\right) $ for every bounded domain $G\subset 
\mathbb{R}^{3}$ \cite{GT:1983}. In particular, this implies that 
\begin{equation}
w\left( \mathbf{x},s\right) =w_{0}\left( z,s\right)
-s^{2}\int\limits_{\Omega }\frac{\exp \left( -s\left\vert \mathbf{x}-\mathbf{%
\xi }\right\vert \right) }{4\pi \left\vert \mathbf{x}-\mathbf{\xi }%
\right\vert }\left[ \epsilon \left( \mathbf{\xi }\right) -1\right] w\left( 
\mathbf{\xi },s\right) d\mathbf{\xi .}  \label{4.17}
\end{equation}%
It is clear from (\ref{4.17}) that the function $\widehat{w}\left( \mathbf{x}%
,s\right) \in C^{\infty }\left( \mathbb{R}^{3}\diagdown \widetilde{\Omega }%
\right) $ for any bounded domain $\widetilde{\Omega }$ such that $\Omega
\subset \widetilde{\Omega },\ \partial \Omega \cap \partial \widetilde{%
\Omega }=\varnothing .$ It is also clear that this function decays
exponentially together with its derivatives as $\left\vert \mathbf{x}%
\right\vert \rightarrow \infty .$ Hence, \ $\widehat{w}\left( \mathbf{x}%
,s\right) \in C^{2+\alpha }\left( \mathbb{R}^{3}\right) .$ Thus, we have
established that the function $w$ satisfies conditions (\ref{1}), (\ref{2}).

We now prove uniqueness of the solution of the problem (\ref{eq:w})--(\ref{2}%
). Assume that there exist two solutions of this problem, $w_{1}$ and $%
w_{2}. $ Then both of them can be represented via (\ref{4.17}). Hence, $%
w_{1}-w_{0}:=\widehat{w}_{1}\in H^{1}\left( \mathbb{R}^{3}\right) $ and $%
w_{2}-w_{0}=\widehat{w}_{2}\in H^{1}\left( \mathbb{R}^{3}\right) .$
Furthermore, both functions $\widehat{w}_{1},\widehat{w}_{2}$ decay
exponentially together with their derivatives as $\left\vert \mathbf{x}%
\right\vert \rightarrow \infty .$ On the other hand, since both these
functions satisfy conditions (\ref{4.15}), (\ref{4.16}), then subtracting
equation (\ref{4.15}) for $\widehat{w}_{2}$ from the same equation for $%
\widehat{w}_{1}$, multiplying the resulting equation by $\widehat{w}_{1}-%
\widehat{w}_{2},$ integrating over $\mathbb{R}^{3}$ and using integration by
parts, we obtain in a standard manner that $\widehat{w}_{1}-\widehat{w}%
_{2}\equiv 0.$ Thus, uniqueness is established.

We now prove estimates (\ref{4.3}). For an arbitrary number $R>0$ let $%
B_{R}=\left\{ \left\vert \mathbf{x}\right\vert <R\right\} .$ Since the
function $\left( \epsilon \left( \mathbf{x}\right) -1\right) w_{0}\left(
z,s\right) \geq 0,$ then the maximum principle applied to (\ref{4.15})
implies that if $\max_{\overline{B}_{R}}\widehat{w}\left( \mathbf{x}%
,s\right) >0,$ then this maximum is achieved at a point $\mathbf{x}%
_{0}\left( R\right) \in \partial B_{R}.$ Hence, setting $R\rightarrow \infty 
$ and using (\ref{4.16}), we obtain $\widehat{w}\left( \mathbf{x},s\right)
\leq 0,$ which proves the right inequality (\ref{4.3}).

We now prove the left inequality (\ref{4.3}). Let 
\begin{equation*}
v_{u}\left( z,t\right) =\frac{\sqrt{\epsilon _{u}}}{2\sqrt{\pi t}}\exp \left[
-\frac{\epsilon _{u}\left( z-z_{0}\right) ^{2}}{4t}\right] 
\end{equation*}%
be the solution of the problem (\ref{4.4})--(\ref{4.5}) for $\epsilon \left( 
\mathbf{x}\right) \equiv \epsilon _{u}.$ Let 
\begin{equation}
p\left( \mathbf{x},t\right) =\int\limits_{0}^{t}\left( v\left( \mathbf{x}%
,\tau \right) -v_{u}\left( z,\tau \right) \right) d\tau .  \label{4.18}
\end{equation}%
Then $\epsilon \left( \mathbf{x}\right) p_{t}-\Delta p=\left( \epsilon
\left( \mathbf{x}\right) -\epsilon _{u}\right) v_{u}$ and $p\left( \mathbf{x}%
,0\right) =0.$ Since $\left( \epsilon \left( \mathbf{x}\right) -\epsilon
_{u}\right) v_{u}>0$ for $t>0$, then the function $p\left( \mathbf{x}%
,t\right) $ is not identical zero. Hence, the maximum principle of Theorem 1
of Chapter 2 of \cite{Friedman:1964} implies that $p\left( \mathbf{x}%
,t\right) >0$ for $t>0.$ Hence, $\left( \mathcal{L}_{2}p\right) \left( 
\mathbf{x},s\right) >0.$ On the other hand, 
\begin{equation*}
\left( \mathcal{L}_{2}p\right) \left( \mathbf{x},s\right) =\frac{1}{s^{2}}%
\left[ \left( \mathcal{L}_{2}v\right) \left( \mathbf{x},s\right) -\left( 
\mathcal{L}_{2}v_{u}\right) \left( z,s\right) \right] =\frac{1}{s^{2}}\left[
w\left( \mathbf{x},s\right) -w_{u}\left( z,s\right) \right] >0.\text{ }%
\square 
\end{equation*}%
\hfill 

The formulation of the global convergence Theorem \ref{th:2} is similar to
the formulation of Theorem 5.1 in \cite{B-K:JIIP2012}. First, the function $%
q^{\ast }\left( \mathbf{x},s\right) \in C^{2+\alpha }\left( \overline{\Omega 
}\right) $ is introduced as well as its Dirichlet boundary condition $\psi
^{\ast }\left( \mathbf{x},s\right) =q^{\ast }\left( \mathbf{x},s\right) \mid
_{\partial \Omega },$ where $q^{\ast }\left( \mathbf{x},s\right) \in
C^{2+\alpha }\left( \overline{\Omega }\right) ,\ \psi ^{\ast }\left( \mathbf{%
x},s\right) \in C^{2+\alpha }\left( \partial \Omega \right) ,\ \forall s\in %
\left[ \underline{s},\overline{s}\right] .$ These functions correspond to
the exact coefficient $\epsilon ^{\ast }\left( \mathbf{x}\right) .$ We also
assume that the function $\psi \left( \mathbf{x},s\right) \in C^{2+\alpha
}\left( \partial \Omega \right) ,\ \forall s\in \left[ \underline{s},%
\overline{s}\right] .$ Next, similarly to functions $q_{n}\left( \mathbf{x}%
\right) $ and $\psi _{n}\left( \mathbf{x}\right) $, functions $q_{n}^{\ast
}\left( \mathbf{x}\right) \in C^{2+\alpha }\left( \overline{\Omega }\right) $
and $\psi _{n}^{\ast }\left( \mathbf{x}\right) \in C^{2+\alpha }\left(
\partial \Omega \right) $ are obtained from functions $q^{\ast }\left( 
\mathbf{x},s\right) $ and $\psi ^{\ast }\left( \mathbf{x},s\right) $,
respectively. The functions $q_{n}^{\ast }$ are solutions of analogs for
equations (\ref{eq:q5}) with Dirichlet boundary conditions $\psi _{n}^{\ast
}.$ There exists a constant $C^{\ast }=C^{\ast }\left( \epsilon
_{u},d\right) >1$ such that 
\begin{eqnarray}
&&\left\Vert q^{\ast }\left( \mathbf{x},s\right) -q_{n}^{\ast }\left( 
\mathbf{x}\right) \right\Vert _{C^{2+\alpha }\left( \overline{\Omega }%
\right) }+\left\Vert \psi ^{\ast }\left( \mathbf{x},s\right) -\psi
_{n}^{\ast }\left( \mathbf{x}\right) \right\Vert _{C^{2+\alpha }\left(
\partial \Omega \right) }  \label{4.200} \\
&\leq &C^{\ast }h,\text{ }s\in (s_{n},s_{n-1}],\ n=1,\dots ,N.  \notag
\end{eqnarray}%
Let $\sigma \in \left( 0,1\right) $ be the level of the error in the
boundary data $\psi \left( \mathbf{x},s\right) $, 
\begin{equation}
\left\Vert \psi \left( \mathbf{x},s\right) -\psi ^{\ast }\left( \mathbf{x}%
,s\right) \right\Vert _{C^{2+\alpha }\left( \partial \Omega \right) }\leq
C^{\ast }\sigma ,\text{ }\forall s\in \left[ \underline{s},\overline{s}%
\right] .  \label{4.20}
\end{equation}%
Introduce the error parameter $\eta =h+\sigma .$ Then (\ref{eq:q7}), (\ref%
{4.200}) and (\ref{4.20}) imply that it is natural to assume that 
\begin{equation}
\left\Vert \psi _{n}-\psi _{n}^{\ast }\right\Vert _{C^{2+\alpha }\left(
\partial \Omega \right) }\leq C^{\ast }\eta ,\text{ }s\in (s_{n},s_{n-1}],%
\text{ }n=1,\dots ,N.  \label{4.21}
\end{equation}

\begin{theorem}[Global convergence]\label{th:2}
 Assume that in (\ref%
{eq:fp1}) $f\left( t\right) =\delta \left( t\right) ,$ the
approximation (\ref{eq:model2}) holds and that the initial tail function $%
V_{1,1}\left( \mathbf{x}\right) $ is calculated as in section \ref{sec:gc}.
Suppose that the maximal number of iterations with respect to $i$ is $m\geq
1 $ and that the algorithm is stopped at a certain $n=\overline{N}\in \left[
1,N\right] .$ Also, let (\ref{4.1}), (\ref{4.20}) and (\ref{4.21}) hold. In
addition, assume that all functions $\epsilon _{n,i}$ satisfy $\epsilon
_{n,i}\left( \mathbf{x}\right) \geq 1$ and that $\left\Vert \epsilon ^{\ast
}\right\Vert _{C^{\alpha }\left( \mathbb{R}^{3}\right) }\leq d/2$. Also, let 
$\underline{s}\geq \underline{s}\left( d\right) $ and $\overline{s}>1.$ Then
there exists a number $B=B\left( \Omega ,\chi ,\overline{s},\alpha ,\epsilon
_{u},d\right) >1$ depending only on listed parameters such that if the error
parameter $\eta $ is so small that 
\begin{equation*}
\eta \in \left( 0,\eta _{0}\right) ,\text{ }\eta _{0}=\frac{1}{B^{6Nm}},
\end{equation*}%
then for $i\in \left[ 1,m\right] ,\ n\in \left[ 1,\overline{N}\right] $ 
\begin{equation}
\left\Vert \epsilon _{n,i}-\epsilon ^{\ast }\right\Vert _{C^{\alpha }\left( 
\overline{\Omega }\right) }\leq B^{3\left( i+\left( n-1\right) m\right)
}\eta \leq \sqrt{\eta }.  \label{eq:error}
\end{equation}%
\end{theorem}

%

\textbf{Outline of the proof}. This proof is similar to the proofs of
Theorem 2.9.4 in \cite{B-K:2012} and Theorem 5.1 in \cite{B-K:JIIP2012}.
Denote by $P_{n,i}=\left\Vert \nabla V_{n,i}-\nabla V^{\ast }\right\Vert
_{C^{1+\alpha }\left( \overline{\Omega }\right) },Q_{n,i}=\left\Vert
q_{n,i}-q_{n}^{\ast }\right\Vert _{C^{2+\alpha }\left( \overline{\Omega }%
\right) },R_{n,i}=\left\Vert \epsilon _{n,i}-\epsilon ^{\ast }\right\Vert
_{C^{\alpha }\left( \overline{\Omega }\right) }.$ First, $P_{1,1}$ is
estimated using (\ref{eq:model2})--(\ref{3.45}) and Schauder theorem \cite%
{GT:1983,L-S-U:1968}. Next, $Q_{1,1}$ is estimated using Schauder theorem
and (\ref{4.21}).\ Next, the number $R_{1,1}$ is estimated using estimates
for $P_{1,1}$ and $Q_{1,1}$ as well as equation (\ref{eq:c}): recall that
functions $\epsilon _{n,i}$ are reconstructed via (\ref{eq:c}). On each
follow up iterative step $\left( n,i\right) $ we first estimate $P_{n,i}.$
Next, we use Schauder theorem to estimate $Q_{n,i}.$ And finally we use (\ref%
{eq:c}) to estimate $R_{n,i}.$

We now outline an idea of estimating numbers $P_{n,i}.$ Using (\ref{eq:tail}%
), we obtain that for $\left( n,i\right) \neq \left( 1,1\right) $ 
\begin{equation}
\nabla V_{n,i}\left( \mathbf{x}\right) -\nabla V^{\ast }\left( \mathbf{x}%
\right) =\frac{\left[ \left( \nabla w_{n,i}-\nabla w^{\ast }\right) w^{\ast }%
\right] \left( \mathbf{x},\overline{s}\right) +\left[ \left( w^{\ast
}-w_{n,i}\right) \nabla w^{\ast }\right] \left( \mathbf{x},\overline{s}%
\right) }{\left( w_{n,i}w^{\ast }\right) \left( \mathbf{x},\overline{s}%
\right) }.  \label{4.23}
\end{equation}%
A similar formula is also valid for $\Delta V_{n,i}\left( \mathbf{x}\right)
-\Delta V^{\ast }\left( \mathbf{x}\right) .$ Hence, using the left estimate (%
\ref{4.3}), we obtain that in (\ref{4.23}) $0<1/\left( w_{n,i}w^{\ast
}\right) \left( \mathbf{x},\overline{s}\right) \leq B_{1}$ with a certain
number $B_{1}>0$ depending on the same parameters as the number $B$. Next,
using the right estimate (\ref{4.3}) as well as arguments of Theorem 2.9.1.2
of \cite{B-K:2012}, we obtain from (\ref{4.23}) an estimate for $P_{n,i}$
via $R_{n,i-1}$. \hfill $\square $

\begin{remark}
\label{re:1} Theorem \ref{th:2} states that if the total number of
iterations with respect to both parameters $n,i$ does not exceed $Nm$ and if
the error parameter $\eta $ is sufficiently small, then all functions $%
\epsilon _{n,i}$ are located in a sufficiently small neighborhood of the
exact solution $\epsilon ^{\ast }.$ The size of this neighborhood is defined
only by the error parameter, which is natural. This property holds
regardless on any a priori knowledge of that small neighborhood. Therefore
we have global convergence within the framework of the approximate
mathematical model of section \ref{sec:gc}.

From the computational standpoint, even though the error estimate (\ref%
{eq:error}) is valid at the first iteration, our numerical observations
suggest that the algorithm should be continued after the first iteration.
The reason is that (\ref{eq:error}) is valid only for the approximate
mathematical model (\ref{eq:model2}). As to which of functions $\epsilon
_{n,i}$ one should consider as the final solution of our inverse problem, it
should be decided on the basis of a stopping criterion, which is chosen
computationally. It is important that we use the same stopping criterion for
all our numerical tests described below. We also note that it is a well
known phenomenon in the field of Ill-Posed Problems that the iteration
number is chosen as one of regularization parameters, see, e.g. pages 156,
157 of \cite{E-H-N:1996}. The approximate mathematical model (\ref{eq:model2}%
) should be verified computationally. This was done in above cited works of
our group as well as in the current paper.
\end{remark}

\section{Data acquisition and preprocessing}

\label{sec:exp}

\subsection{Data acquisition}

\begin{figure}[tph]
\centering
\subfloat[]{\includegraphics[width=0.45\textwidth,height=0.35%
\textwidth]{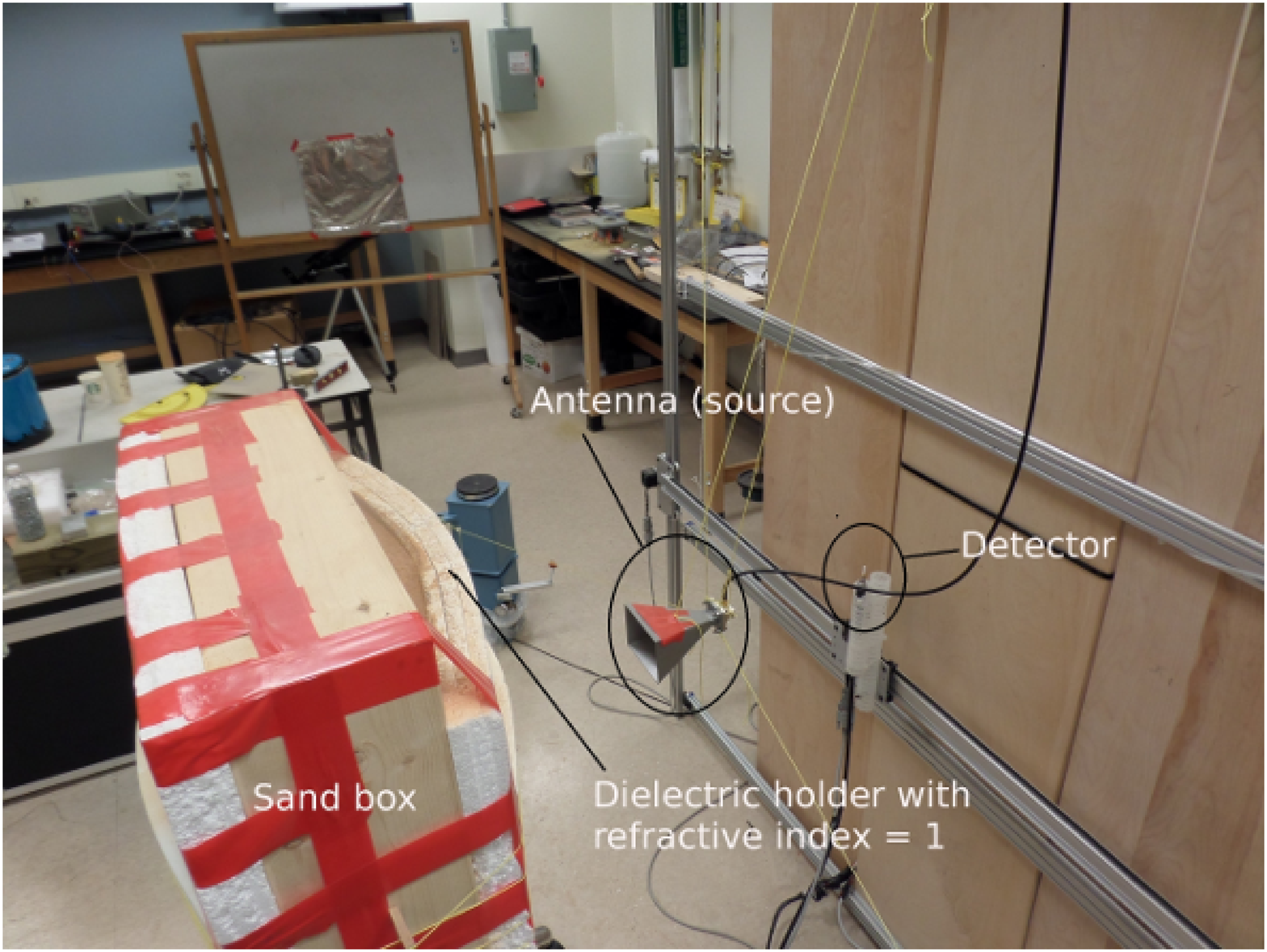}} \hspace{0.3truecm} \subfloat[]{%
\includegraphics[width=0.4\textwidth,height=0.35%
\textwidth]{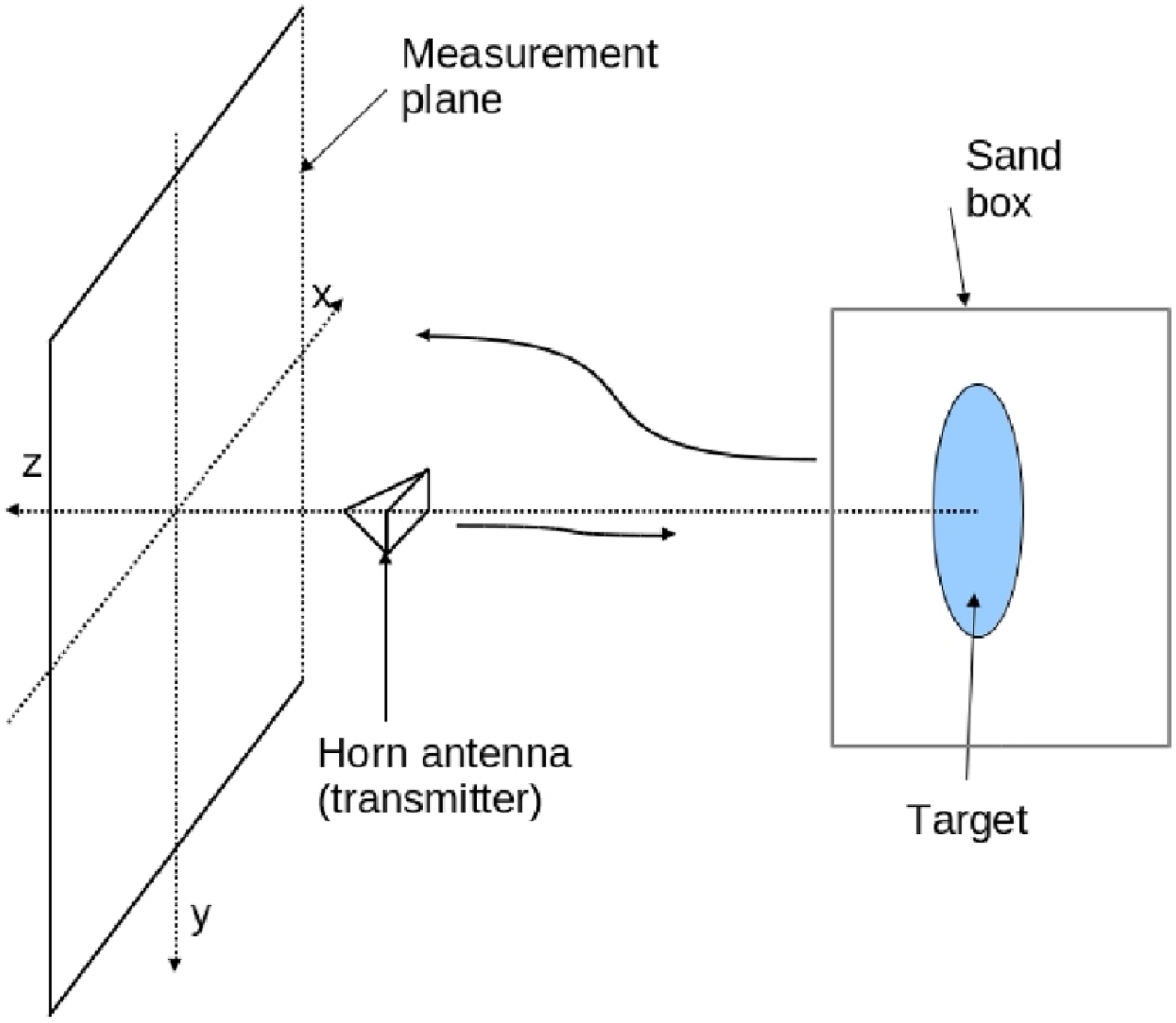}}
\caption{(a): Experimental setup; (b) Schematic diagram of our data
acquisition.}
\label{fig:setup}
\end{figure}

The collection of the experimental data was carried out using the same
configuration as in \cite{BTKF:IP2013,TBKF:SISC2014}, where we imaged
targets placed in air, see the detailed description in section 3 of \cite%
{TBKF:SISC2014}. The only difference is that in the current paper the
objects considered were placed inside a box filled with dry sand with the
dielectric constant of $\epsilon \left( \text{sand}\right) =4$. This was
used to model the case of buried objects, see Figure~\ref{fig:setup}. 

The sand box was placed in front of the transmitter. Its front and back
sides were covered by Styrofoam whose dielectric constant is approximately
1, i.e., it does not affect the incident and scattered signals. The
transmitter (a horn antenna) was fixed at a given position while a detector
was scanned over a square of a vertical plane behind the transmitter, which
we refer to as the measurement plane. At each detector location, the source
emitted an electric pulse of 300 picoseconds (ps) duration, and the
time-dependent scattered wave was captured by the detector. Then the
detector was moved to the next location and the measurement was repeated.
Hence, our data can be considered as generated by a single point source.
Consider the Cartesian coordinate system $0xyz$ as shown in Figure \ref%
{fig:setup}(b). Below \textquotedblleft m" stands for meter. The detector
was scanned in the square of 1 m by 1 m with the step size of 0.02 m,
starting at $(x,y)=(-0.5,-0.5)$ and ending at $(x,y)=(0.5,0.5)$. The horn
antenna was placed at the distance of about 0.2--0.25 m from the measurement
plane, and the distance from the sand box to the measurement plane was about
0.7--0.8 m. 

The wavelength of the incident pulses was about 0.04 m. The time step
between two consecutive records was $\Delta t=10$ ps. Each signal was
recorded for 10 nanoseconds. Since the source was located far away (at about
12--15 wavelengths) from the targets, the use of the incident plane wave in
our model (\ref{eq:fp1})--(\ref{eq:fp2}) is well suited.

\subsection{Data preprocessing}

As we have mentioned in Introduction, there is a \textit{huge misfit}
between our experimental data and the data produced via computational
simulations. We have pointed out in \cite{TBKF:SISC2014} that there are
several causes of this misfit such as: (i) the instability of the amplitude
of the emitted signals which causes the instability of the backscatter
signals, (ii) unwanted waves scattered by several existing objects around
our device (see Figure 3.2(a) of \cite{TBKF:SISC2014}), (iii) the shadow on
the measurement plane caused by the horn antenna since it was placed between
the sand box and the measurement plane, see Figures \ref{fig:setup}, and
(iv) the difference between the experimental and simulated incident waves.
Due to some limitations, it was technically very hard to place the horn
antenna behind the measurement plane. Therefore, the \emph{central procedure}
required before applying inversion methods is data preprocessing. This
procedure is inevitably heuristic and cannot be rigorously justified. 
In the case of targets located in air we proposed in \cite{TBKF:SISC2014} a data preprocessing procedure which consists of the
following steps:

1. \textit{Off-set correction}: The recorded signals may be shifted from the
zero mean value. This can be corrected by subtracting the mean value from
them.

2. \textit{Time-zero correction}: Time-zero refers to the moment at which an
incident signal is emitted from the transmitter. This step shifts the data
to the correct time-zero.

3. \textit{Data propagation}: In this step, the data is propagated to a
plane, which we refer to as the \textit{propagated plane}. This plane is
located closer to the targets than the measurement plane. This means that we
approximate the scattered wave on the propagated plane using the measured
scattered wave on the measurement plane. There are two reasons for doing
this \cite{TBKF:SISC2014}. The first one is that since the kernel of the
Laplace transform (\ref{200}) decays exponentially with respect to time,
which is proportional to the distance from the targets to the measurement
plane, then the amplitude of the data after the Laplace transform on the
measurement plane is very small and can be dominated by the computational
error. The second reason is that the data propagation procedure helps to
reduce the computational cost substantially since the size of the
computational domain $\Omega $ is reduced. We have also observed that the
data propagation helps to reduce the noise in the measured data.

4. \textit{Extraction of target's signal}: This is the most difficult step
of data preprocessing. Apart from the signals scattered from the targets,
our measured data also contain various types of unwanted signals and noise.
The unwanted signals which come earlier than the targets' signals dominate
the latter after the Laplace transform. Thus, this step helps to remove both
these unwanted signals and noise. In addition to unwanted signals
and noise as in the case of targets located in air, in our situation of
targets buried in the sand this step also removes the signal caused by the sand, see section \ref{sec:ext}. 

5. \textit{Source shift}: In our simulation, we assume that the incident
plane wave is emitted on the $\{z = z_0\}$. Therefore, the computational
cost depends on the distance from this plane to our targets. To avoid
unnecessary computational cost in our forward and inverse solvers, we
artificially shift the source closer to the sand box. This is done by
shifting the whole time-dependent data in time.

6. \textit{Data calibration}: Finally, since the amplitude of the
experimental signals are usually quite different from the simulated ones, we
scale the former to better match the latter in amplitude by multiplying the
former by a certain factor, which we call the \emph{calibration factor}. The
choice of this factor is based on the data for a known target referred to as
the \textit{calibrating object}.

Steps 1, 2, 5 and 6 are basically the same as in \cite{TBKF:SISC2014}. We
just note that, in this paper, only one calibration factor was used in Step
6 for both non metallic and metallic targets in contrast to the data used in 
\cite{BTKF:IP2013,TBKF:SISC2014} where two different factors were used for
these two types of target. However, steps 3 and 4 used in this paper are
different from those of the above works, therefore we present them below.
Moreover, in Step 4 we also present a method for estimating the burial depth
of a target, which is defined as the distance from the front surface of the
target to the sand's surface.

\subsubsection{Data propagation}

\label{sec:dataprop}


In \cite{TBKF:SISC2014} we proposed a time-reversal data propagation method.
The idea of that method is to solve a time-domain wave equation with the
reversed time variable. In this paper, we make use of another data
propagation method which is based on the data back-propagation via the
Fourier transform. This technique is known as the Stolt migration and it is
popular in Geophysics, see \cite{Stolt:1978,Yilmaz:1987}. However, in the
standard Stolt migration the wave at the initial time is calculated in the
whole spatial domain of interest, whereas we calculate the wave only at a
plane parallel to the measurement plane but in the whole time interval. This
technique is described as follows.

%
%
%
%

We assume that the scattered wave propagates in the positive $z$-direction.
Denote by $P_{m}=\{z=b\}$, $b>0$, the measurement plane and by $P_{p}=\{z=a\}
$, with $a<b$, the propagated plane, which is closer to the target of interest than $%
P_{m}$. We also denote by $u^{s}(\mathbf{x},t)$ the scattered wave. Our
objective here is to determine the function $g(x,y,t):=u^{s}(x,y,a,t),$
given the measured data $f\left( x,y,t\right) :=u^{s}(x,y,b,t)$. We assume
that the medium is homogeneous in the half space $z\geq a$ with $\epsilon
\equiv 1$. Therefore, $u^{s}$ is the solution of the following problem: 
\begin{eqnarray}
&&u_{tt}^{s}-\Delta u^{s}=0,\ \mathbf{x}\in \mathbb{R}^{3},\ z\geq a,\ t\in
(0,\infty ),  \label{eq:dp1} \\
&&u^{s}(\mathbf{x},0)=u_{t}^{s}(\mathbf{x},0)=0.  \label{eq:dp2}
\end{eqnarray}%
Consider the Fourier transform 
\begin{equation}
\hat{u}^{s}(k_{x},k_{y},z,\omega )=\int\limits_{0}^{\infty
}\int\limits_{-\infty }^{\infty }\int\limits_{-\infty }^{\infty
}u^{s}(x,y,z,t)e^{-i(\omega t+xk_{x}+yk_{y})}dxdydt.  \label{eq:dp3}
\end{equation}%
It follows from (\ref{eq:dp1}) that $\hat{u}^{s}$ satisfies the equation: 
\begin{eqnarray}
\hat{u}_{zz}^{s}+(\omega ^{2}-k_{x}^{2}-k_{y}^{2})\hat{u}^{s} &=&0,\ z\geq a,
\label{eq:dp4} \\
\hat{u}(k_{x},k_{y},z,\omega ) &=&\hat{g}(k_{x},k_{y},\omega ),  \label{100}
\end{eqnarray}%
where $\hat{g}(k_{x},k_{y},\omega )$ is the Fourier transform (\ref{eq:dp3})
of $g(x,y,t).$ We consider two cases:

\textbf{Case 1}: $\omega ^{2}-k_{x}^{2}-k_{y}^{2}<0$. Keeping in mind that
the scattered wave propagates in the positive $z$-direction, the problem (%
\ref{eq:dp4})--(\ref{100}) has the following solution 
\begin{equation}  \label{eq:case1}
\hat{u}^{s}_1(k_{x},k_{y},z,\omega )=\hat{g}(k_{x},k_{y},\omega )\exp \left(
-(z-a)\sqrt{k_{x}^{2}+k_{y}^{2}-\omega ^{2}}\right) ,\text{ }z>a.
\end{equation}

\textbf{Case 2}: $\omega ^{2}-k_{x}^{2}-k_{y}^{2}\geq 0$. Then the solution $%
\hat{u}^{s}$ can be represented as 
\begin{equation}  \label{eq:case2}
\hat{u}^{s}_2(k_{x},k_{y},z,\omega )=\hat{g}(k_{x},k_{y},\omega )\exp \left(
-i(z-a)\sqrt{\omega ^{2}-k_{x}^{2}-k_{y}^{2}}\right) ,\text{ }z>a.
\end{equation}%
The negative sign in the exponential term in this formula is due to the fact
that the scattered wave is out-going in the positive $z$-direction.

Since the solution (\ref{eq:case1}) is exponentially decaying as $%
z\rightarrow \infty ,$ which represents the evanescent wave, it practically
cannot propagate to the measurement plane, which is in the far field zone.
Hence, 
\begin{equation*}
\hat{f}(k_{x},k_{y},\omega )=\hat{u}_{2}^{s}(k_{x},k_{y},b,\omega )=\hat{g}%
(k_{x},k_{y},\omega )\exp \left( -i(b-a)\sqrt{\omega ^{2}-k_{x}^{2}-k_{y}^{2}%
}\right) .
\end{equation*}%
Using the inverse Fourier transform, we obtain 
\begin{equation}
g(x,y,t)=\iiint\limits_{\omega ^{2}-k_{x}^{2}-k_{y}^{2}>0}\hat{f}%
(k_{x},k_{y},\omega )e^{i(b-a)\sqrt{\omega ^{2}-k_{x}^{2}-k_{y}^{2}}%
}e^{i(\omega t+xk_{x}+yk_{y})}dk_{x}dk_{y}d\omega .  \label{eq:dp5}
\end{equation}%
Given the data $f(x,y,t)$ at the measurement plane, we compute $\hat{f}$ as
well as $g(x,y,t)$ via (\ref{eq:dp5}) using the Fast Fourier Transform.

For each data set, the propagated plane $P_{p}$ was determined as follows.
We first propagated the data to the sand's surface. Using this propagated
data, we estimated the burial depths of the targets (section \ref{sec:ext}).
Next, if the burial depth of the target closest to the sand surface was
larger than 4 cm, we propagated the data again from the measurement plane up
to the plane $P_{p},$ whose distance to the front surface of that target was
approximately 4 cm. Otherwise, we used the data propagated up to the sand's
surface for the next step of data preprocessing. Note that even we
propagated the data beyond the sand's surface, we still saw the reflection
from the sand's surface in the propagated data since we did not take into
account the presence of the sand box in the data propagation, i.e.,
when propagating the data in the sand, we assumed that $\epsilon =1$
 in the sand. This reflection from the sand's surface was removed when the
targets' signals were extracted, see section \ref{sec:ext}. Note that the
grid points at $P_{p}$ are the same as the ones at the measurement plane $%
P_{m}.$ Thus, below we call \textquotedblleft detectors" the grid points at
the propagated plane $P_{p}$.

\begin{figure}[tph]
\centering
\subfloat[Before propagation]{\includegraphics[width =
0.48\textwidth]{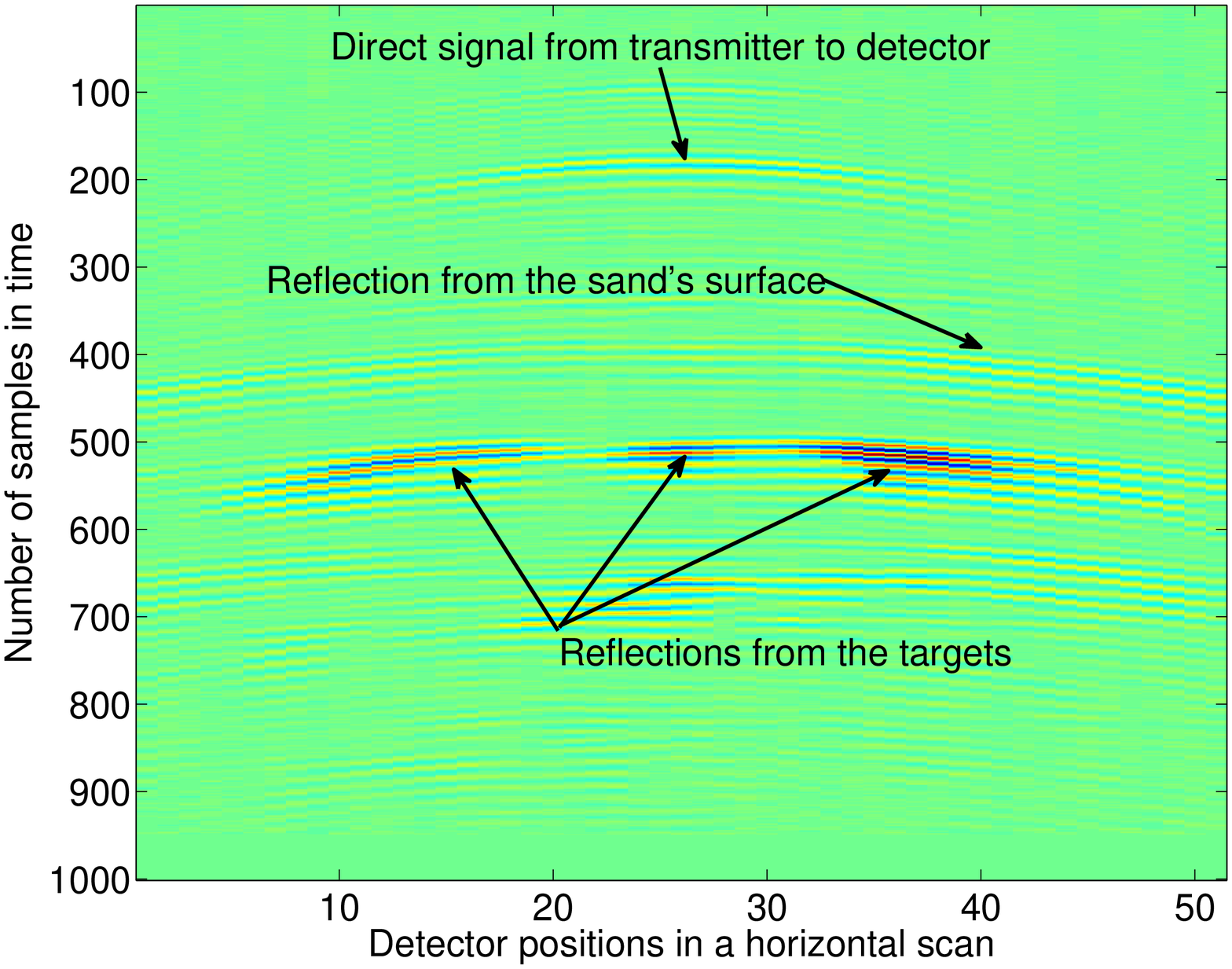}} 
\subfloat[After
propagation]{\includegraphics[width = 0.48\textwidth]{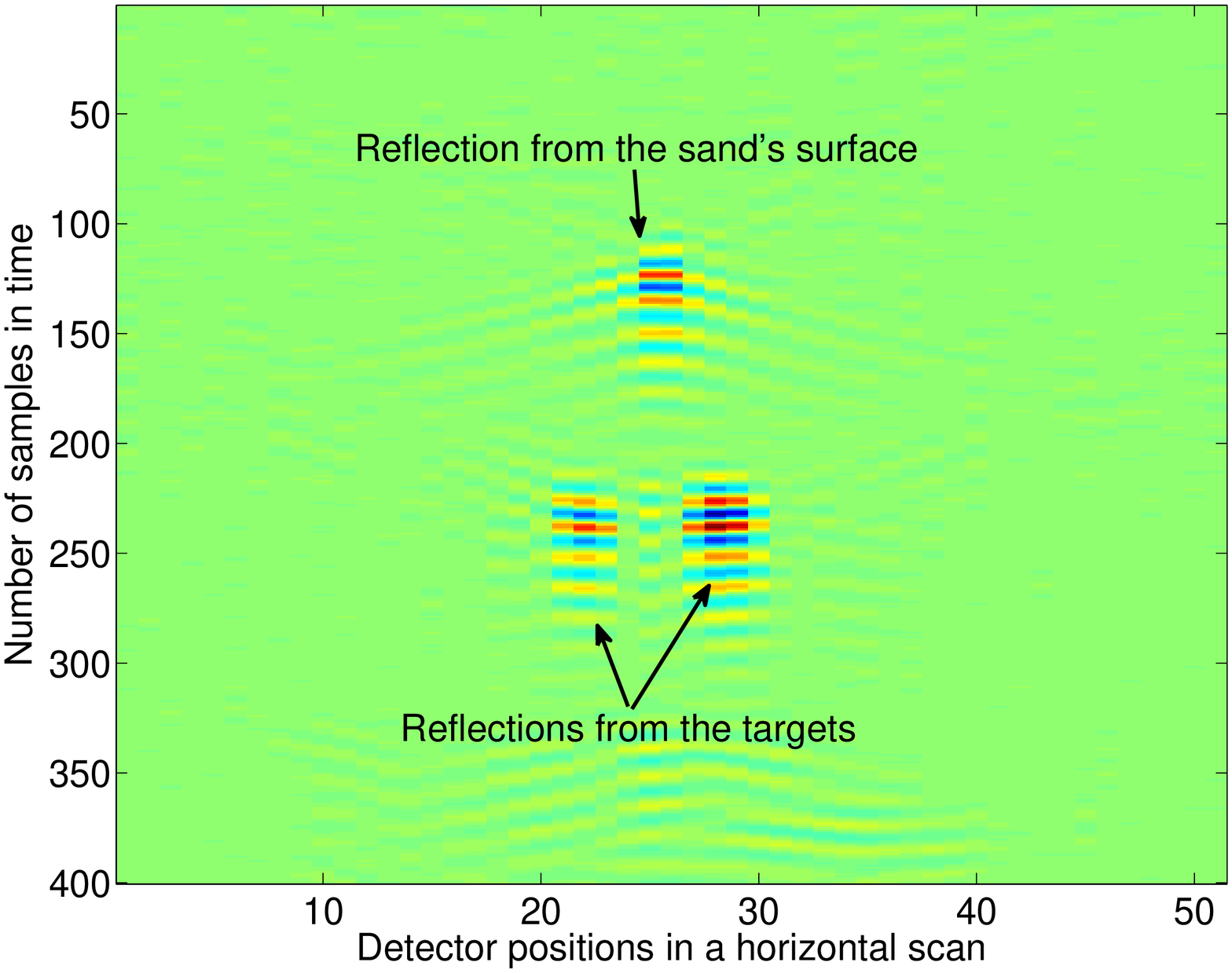}}
\caption{Result of the data propagation for signals from two targets buried
inside the sand box. The signals of the two targets are well separated from
each other as well as from the reflection from the sand's surface after the
data propagation.}
\label{fig:dataprop}
\end{figure}

A result of the data propagation is illustrated in Figure \ref{fig:dataprop}%
. The figure shows a horizontal scan of the sand box containing two buried
metallic targets. The horizontal side denotes the indices of the detector's
locations and the vertical side denotes time. Time increases from the top to
the bottom. The propagation distance, $b-a$, is $0.8$ m, which means that 
the propagated plane $P_{p}$ almost  coincides with the sand's
surface. Figure \ref{fig:dataprop}(a) shows the original data while Figure %
\ref{fig:dataprop}(b) shows the data after the propagation. As can be seen
from these figures, the targets' signals in the original data are smeared
out. On the other hand, they are focused after the data propagation making
the two targets more clearly distinguished. This is well known for migration
methods. Moreover, we can also see that the reflection from the sand's
surface is also more visible after the propagation and is well separated
from the targets' signals.

\subsubsection{Estimation of the burial depth}

\label{sec:ext}

Since the sand's surface reflects our microwave pulses and these reflected
waves arrive at the detectors before the ones reflected from the targets
(see Figure \ref{fig:dataprop}(b)), the targets' signals are dominated by
that of the sand's surface after the Laplace transform. In addition, the
measurement noise appearing earlier than the targets' signals also affects
the latter after the Laplace transform due to the exponential decay of the
kernel, see Figure \ref{fig:4.4}(a). Therefore, an additional important data
preprocessing step is needed in order to pick up only the reflection from
the targets and remove unwanted signals and noise coming earlier than the
targets' signals. This step is applied to the propagated data as described
below. 

We first define some terms which are used in this section. These
terms are related only to the propagated plane $P_{p}.$ The 
\textit{strongest detector} in a data set is defined as the detector at
which the recorded signal has the largest amplitude. A \textit{strong target}
is either a metallic one or a nonmetallic one whose dielectric constant is
larger than that of the sand. If the dielectric constant of a target is
smaller than that of the sand, we call it a \textit{weak target}. The 
\textit{strongest negative (positive) peak} of a time-dependent signal at a
certain detector's location is the negative (positive) peak whose amplitude
is larger than the amplitudes of other negative (positive) peaks of the same
signal.

\begin{figure}[tph]
\centering
\subfloat[A strong target]{\includegraphics[width =
0.48\textwidth]{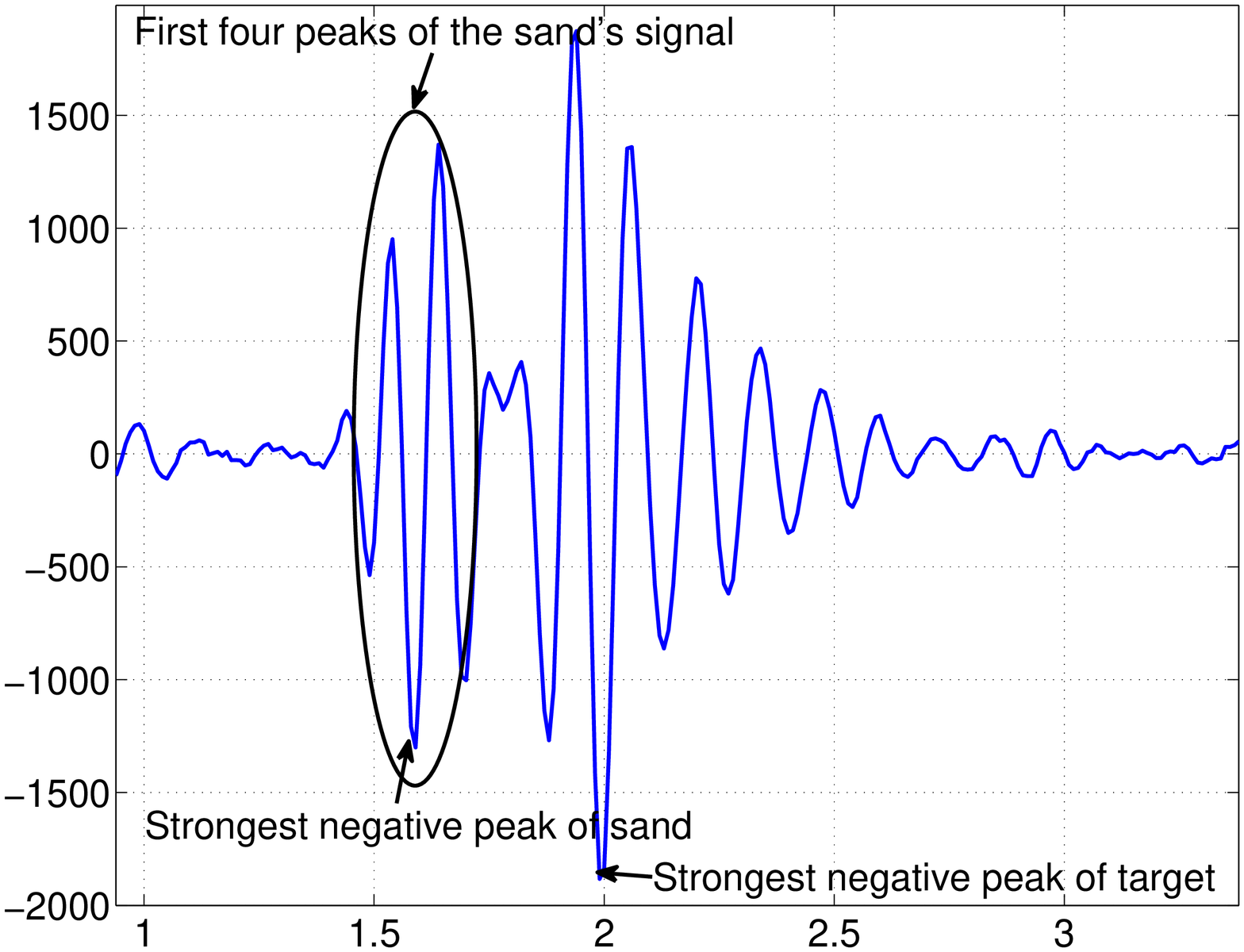}} 
\subfloat[A weak
target]{\includegraphics[width = 0.48\textwidth]{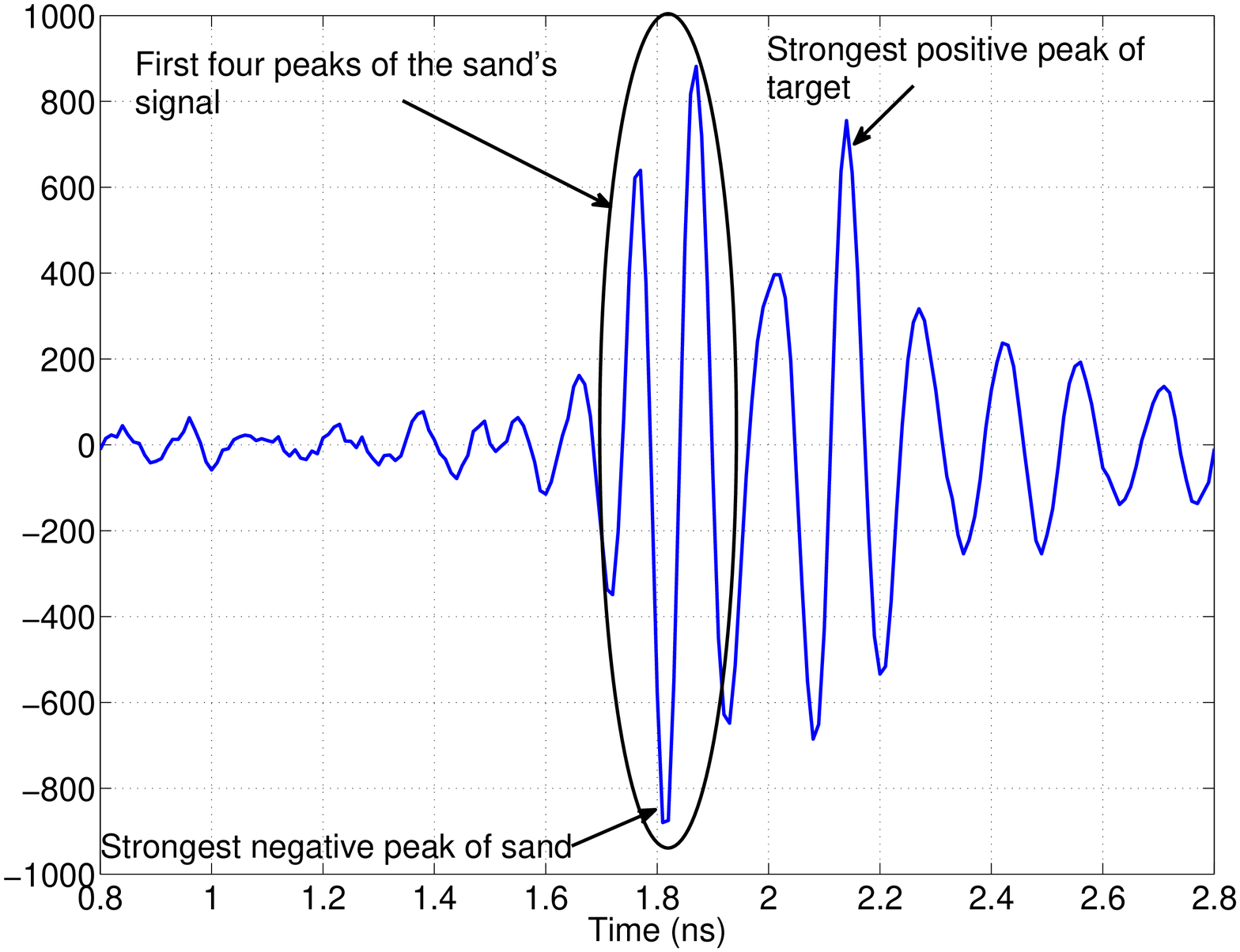}}
\caption{One-dimensional propagated signals at the strongest detectors of
two targets: one strong target and one weak target. The signals consist of
the reflection from the sand's surface followed by the reflections from the
targets.}
\label{fig:4.3}
\end{figure}

We first estimate the burial depth of a target in each data set. For this
purpose, we took the strongest detector. We first determined the strongest
negative peak among the first four peaks, starting from the first negative
peak, see Figure \ref{fig:4.3}. This strongest negative peak is considered
as the strongest negative peak of the sand's signal. After that, we excluded
those first four peaks. The reason for considering those four peaks was due
to our observation that those peaks should belong to the reflection from the
sand's surface. Moreover, the first two negative (so as two positive) peaks
of the incident wave were increasing in amplitude. After that, the negative
(so as positive) peaks of the incident wave decreased in amplitude.
Therefore any increase in amplitude of the peaks after those first four
peaks should be due to the reflection from the target. By detecting the next
negative (or positive) peak which was stronger than the previous negative
(positive) one, we located the target's signal. Then, we determined the
strongest negative peak of the target's signal. Denoting by $\Delta t$ the
time delay between the latter peak and the strongest negative peak of the
sand's signal, the burial depth of the target was approximated by $n(\text{%
sand})\Delta t$, where $n(\text{sand})=\sqrt{\epsilon (\text{sand})}$ is the
refractive index of the sand.

\subsubsection{Extraction of target's signal: the most difficult step of
data preprocessing}

\label{sec:extr}

After estimating the burial depth, we extracted the target's signal. The
extraction of signals of targets in air is quite simple. However, it is 
\emph{very challenging} in the case of buried objects, especially for weak
targets. Indeed, when a weak target is buried at a shallow depth, its signal
is merged with the reflection from the sand's surface. When it is buried at
a deep depth, its signal is usually too weak to be visible in the data. Our
experimental observations have shown that if a weak target is buried at a
depth of more than 5 cm, then we cannot detect it. In this case, the target
is missed.

As in estimating the burial depth, we also worked with the strongest
detector first and excluded the first four peaks. After that, we selected
the target's signal as follows: (i) Suppose that either the burial
depth was larger than 5 cm, or the strongest negative peak of the target's
signal was larger than that of the sand's signal in amplitude. Then
we choose as the first peak of the target's signal the strongest negative
peak located after the excluded ones; (ii) otherwise, the first peak of the
target's signal was determined as the first positive peak which was larger
than the previous positive one, provided that such a peak exists, see Figure %
\ref{fig:4.3}. Since the reconstructed dielectric constants of targets of
case (i) (respectively, case (ii)) was always larger (smaller) than that of the
sand, we also categorized a target in case (i) (case (ii)) as a
strong (weak) target. For all other detectors, we started from those closest
to the strongest detector and on each of them assigned as the first peak of
the target's signal the one which was time wise closest to that of the
strongest detector. For strong targets this one should be a negative peak,
while it was a positive peak for weak targets. Next, we continued similarly
on all other detectors via sequentially choosing those peaks closest to the
one of the previous detector. The reason for choosing a negative (positive)
peak as the first peak of the target's signal for strong (weak) targets was
due to our observations in numerical simulations and experimental data which
have indicated that:

a. For a strong target, the first peak of the target's signal should be
negative.

b. For a weak target, the first peak of the target's signal should be
positive.

Moreover, if a strong target is buried at a depth less than 5 cm, then its
signal is stronger than that of the sand in amplitude. If the burial depth
is more than 5 cm, then its signal might not be stronger than that of the
sand. However, since, as we mentioned above, weak targets are not visible at
depths larger than 5 cm, we consider all targets buried at these depths as
strong targets.

In all above cases, the data before the chosen first peak of the target's
signal were set to zero. Hence, the Laplace transform of the preprocessed
data is not affected by values before the first chosen peak. We note that
such a choice of starting peaks artificially immerses our targets in air:
because we exclude the reflection from the sand's surface. Therefore, what
we reconstruct for each target by the globally convergent method is the
ratio between its dielectric constant (or the \textit{effective} dielectric
constant for metals) and that of the sand, $\epsilon \left( \text{target}%
\right) /\epsilon \left( \text{sand}\right) $. Next, to obtain the value of
the dielectric constant of the target, we multiply this ratio by $\epsilon
\left( \text{sand}\right) =4$.

\begin{figure}[tph]
\centering
\par
\subfloat[Before the extraction of the targets'
signals]{\includegraphics[width = 0.52\textwidth]{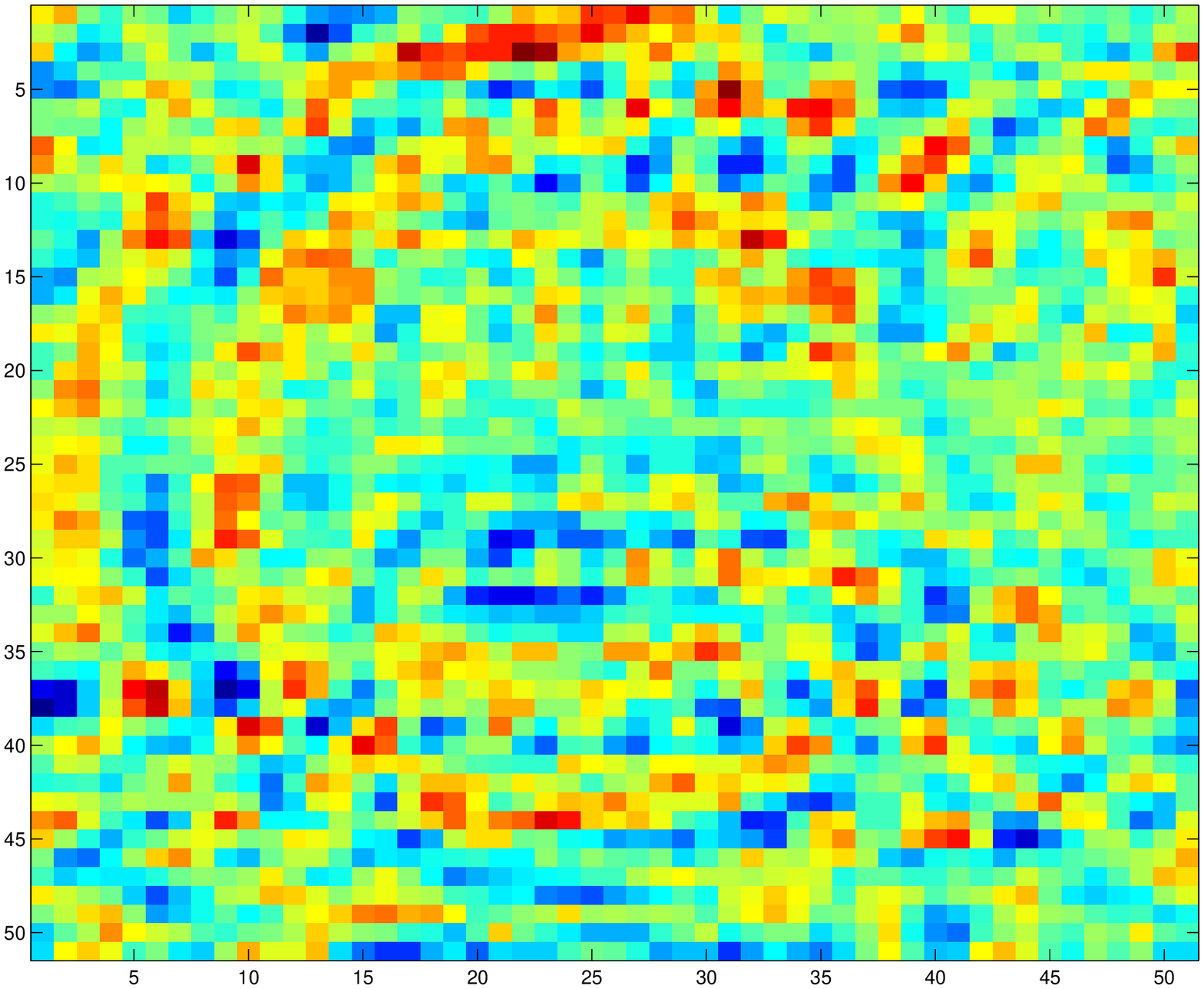}} 
\subfloat[After the extraction of the targets'
signals]{\includegraphics[width = 0.48\textwidth]{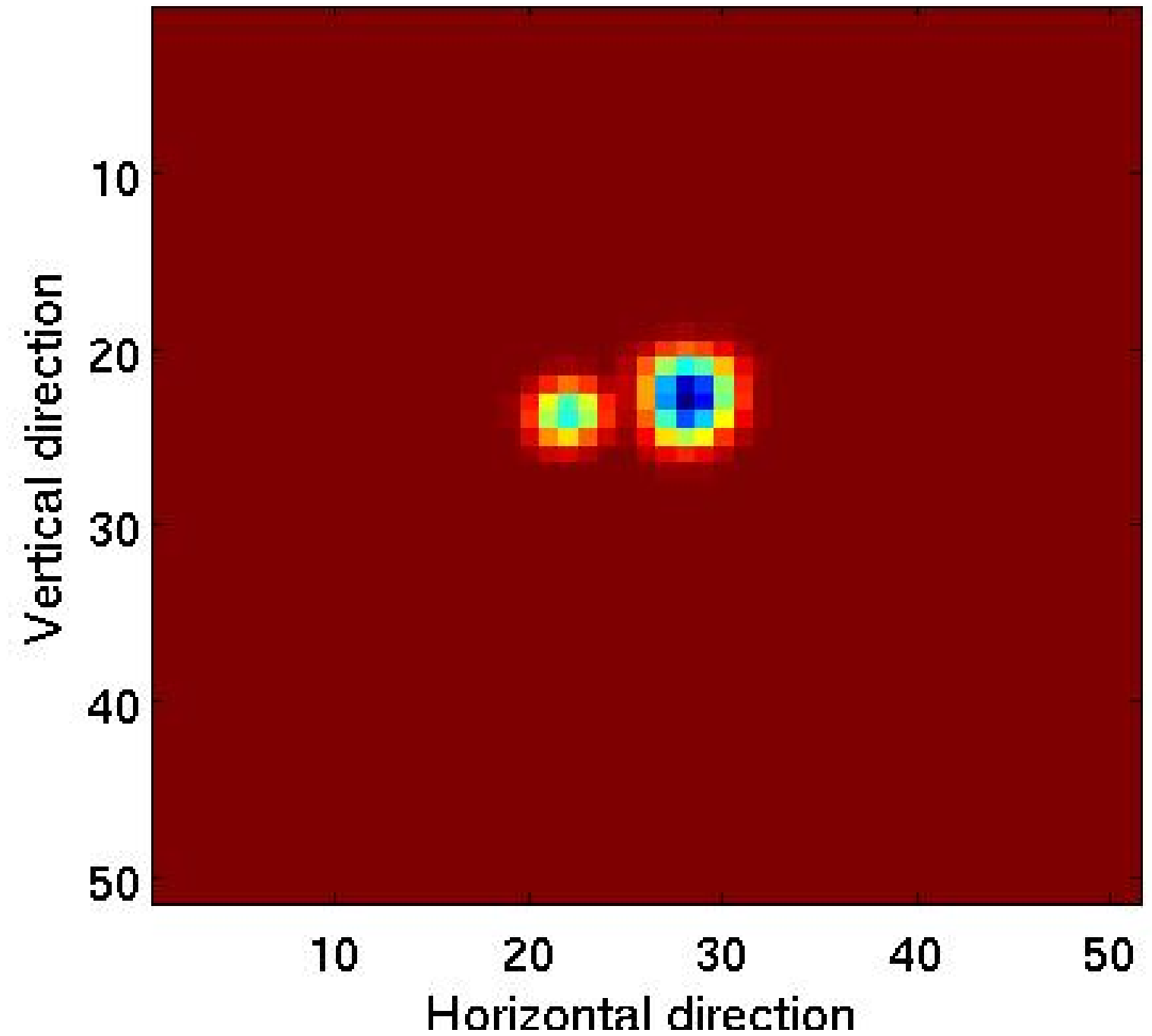}}
\caption{The Laplace transforms of the data on the propagated plane before
(a) and after (b) the extraction of the targets' signals. Without the
extraction, we cannot see the targets. After the extraction, the two targets
show up clearly.}
\label{fig:4.4}
\end{figure}

Figure \ref{fig:4.3} shows one-dimensional propagated signals at the
strongest detectors of a strong target and a weak target. We indicate
there the sand's signal and the peaks of the targets. These peaks were
chosen as the first peaks of signals from the targets. Samples of the
Laplace transform of the data before and after the extraction of the
targets' signals are shown in Figure \ref{fig:4.4}, which indicate the
necessity of this preprocessing step.

Figure \ref{fig:4.4} (b) also shows that the preprocessed data allow
us to estimate locations of the targets in $x,y$ directions as
well as their $xy$-cross sections, see Section 4.3 of \cite{TBKF:SISC2014}
for the method we proposed for estimating the cross sections of targets.
These types of information help to reduce the domain in which we look for
the targets. Indeed, in Test 2 below, we took into account these types of
information in choosing the first tail function. 

\section{Reconstruction results}

\label{sec:num}

We now illustrate the performance of the globally convergent algorithm of
Section \ref{sec:gca} for our experimental data sets.

\subsection{Description of experimental data sets}

To have a rather systematic study of the performance of the globally
convergent algorithm for these data, we have tested this algorithm for
different types of targets at different burial depths varying between 0.02 m
and 0.14 m. Our results in this paper include 25 data sets. Table \ref{ta:1}
describes the details of these data sets including the targets' materials.
Among them, there are 10 non-blind cases and 15 blind ones.
\textquotedblleft Blind" means that the targets were unknown to the
computational team (NTT, LB, MVK) but known to MAF, who was leading the data
collection process. The non-blind targets are the ones that were known to
NTT and MVK but were not known to LB who performed Test 1 below. 
Note that we were able to directly measure refractive indices $n=\sqrt{%
\epsilon }$ rather than dielectric constants. The refractive
indices of all non metallic targets were measured after the reconstruction
results were obtained by the computational team. Next, computational results
were compared with directly measured ones. Some of the non-blind targets
were used for calibrating and fine tuning of the reconstruction procedure.
The blind targets were used to ensure that this procedure works in realistic
blind data cases. It is important that the same reconstruction
procedure, with the same choice of parameters, was used for all targets.

\begin{table}[tbp]
\caption{{\protect\small \emph{Description of the test data sets. Seven of
them consist of two targets each (5, 6, 16, 17, 20, 23, 25). Two targets can
be considered as heterogeneous (11, 12).}}}
\label{ta:1}
\begin{center}
{\footnotesize \ 
\begin{tabular}{|c|l|l|l|}
\hline
Object & Blind/ & Description of target & Material \\ 
\# & Non-blind &  &  \\ \hline
1 & Non-blind & A metallic cylinder & Metal \\ \hline
2 & Non-blind & A metallic ball & Metal \\ \hline
3 & Non-blind & A bottle filled with clear water & Water \\ \hline
4 & Non-blind & A wet wooden block & Wet wood \\ \hline
5 & Non-blind & Two metallic blocks at 6 cm separation & Metal/Metal \\ 
\hline
6 & Non-blind & A metallic cylinder and a teflon bar & Metal/Teflon \\ \hline
7 & Non-blind & A metallic block & Metal \\ \hline
8 & Non-blind & An empty bottle & Air \\ \hline
9 & Non-blind & A bottle filled with teflon bars & Teflon \\ \hline
10 & Blind & A ceramic mug & Ceramic \\ \hline
11 & Blind & A wooden doll filled with metallic screws & Wood/Metal \\ 
&  & (heterogeneous, diffuse scattering) &  \\ \hline
12 & Blind & A geode (heterogeneous): & Rock \\ 
&  & two spherical layers and air inside &  \\ \hline
13 & Blind & A piece of rock & Rock \\ \hline
14 & Blind & A plastic bottle filled with coffee grounds & Coffee grounds \\ 
\hline
15 & Blind & A ceramic mug & Ceramic \\ \hline
16 & Blind & A cylinder and a block at 3 cm separation & Metal/Metal \\ 
\hline
17 & Blind & An aluminum can and a block & Metal/Metal \\ \hline
18 & Blind & A wooden doll with a metallic block inside & Wood/Metal \\ 
&  & (heterogeneous) &  \\ \hline
19 & Blind & A bottle of water & Water \\ \hline
20 & Blind & A metallic block and a rock & Metal/rock \\ \hline
21 & Blind & A steel mug & Metal \\ \hline
22 & Blind & A wet wooden block & Wet wood \\ \hline
23 & Blind & A wet wooden block and an empty bottle & Wet wood/air \\ \hline
24 & Blind & A wet wooden block & Wet wood \\ \hline
25 & Non-blind & Two metallic blocks at 1 cm separation & Metal/Metal \\ 
\hline
\end{tabular}
}
\end{center}
\end{table}

The burial depths of the targets varied between 2.5 centimeters (cm) to 14
cm. Note that typically burial depths of antipersonnel land mines do not
exceed 10 cm. Among our targets were metallic blocks, wooden
blocks, metallic cylinders, metallic spheres, wooden dolls, etc. Seven data
sets (\#5, 6, 16, 17, 20 23, 25) consisted of two targets each. In
particular, there were three targets (\#11, 12, 18) which can be considered
as heterogeneous ones, see Table \ref{ta:1}. The data of the sand box alone
(without buried objects) was used for the calibration of our data.

\subsection{Numerical implementation}

\label{sec:num1}

Details of the numerical implementation of the globally convergent algorithm
can be found in \cite{B-K:2012, TBKF:SISC2014}. For the reader's
convenience, we briefly describe the main points here.


\textbf{Choosing the domain and solving the forward problem:} In the
preprocessed data, the distance from the front sides of the targets to the
backscattering boundary of our inversion domain $\Omega $ was $0.04$ m. The
reason for choosing this distance was due to good reconstruction results we
had obtained for several non-blind targets. Hence, the inversion domain $%
\Omega $ was chosen as 
\begin{equation}
\Omega =\left\{ \mathbf{x}\in (-0.4,0.4)\times (-0.4,0.4)\times
(-0.2,0.04)\right\} .  \label{Y}
\end{equation}%
Moreover, since it is impossible to solve the problem (\ref{eq:fp1})--(\ref%
{eq:fp2}) in the entire space, in numerical computation, we approximated it
by an initial boundary value problem in a bounded domain $G\subset \mathbb{R}%
^{3}$ such that $\Omega \subset G$, see \cite{B-K:2012,TBKF:SISC2014} for
details. In this paper, we choose $G$ as the rectangular parallelepiped 
\begin{equation}
G=\left\{ \mathbf{x}\in (-0.5,0.5)\times (-0.5,0.5)\times (-0.3,0.3)\right\}
.  \notag
\end{equation}%
This domain $G$ was decomposed into two subdomain: $G=\Omega \cup
(G\setminus \Omega )$. We recall that $\epsilon (\mathbf{x})=1$ in $%
G\setminus \Omega $. Therefore, it is only necessary to solve the inverse
problem in $\Omega $. A finite element mesh with tetrahedral elements is
used in $\Omega $, while in $G\setminus \Omega $ we use a finite difference
mesh with mesh sizes of $0.02\times 0.02\times 0.02$. The forward problem
was solved using the software package WavES \cite{waves} via a hybrid finite
difference/finite element method described in \cite{BSA}.

The time interval on which the problem (\ref{eq:fp1})--(\ref{eq:fp2}) was
solved was chosen to be $\left( 0,T\right) =(0,1.2)$ since no relevant signals appeared after this time interval. Note that the time
variable was scaled so that the dielectric constant of air is equal to $1$. Since the explicit
scheme in time was used in WaveES, the time step size was chosen as $\Delta
t=0.0015$ which satisfied the CFL stability condition. 

The pseudo frequencies $s_{n}$ were chosen from $\underline{s}=7$ to $%
\overline{s}=9$ with the step size $h=0.05$. This pseudo frequency interval
was chosen because it gave good reconstructions of the non-blind targets. We
have observed that the pseudo frequency interval $s\in \lbrack 8,10]$ also
provided good reconstruction results.

In our simulations, the waveform function $f$ in (\ref{eq:fp1}) was chosen
by $f(t)= 2\omega \cos (\omega t)$ for $0\leq t\leq t_{1}=2\pi /\omega,$ and 
$f(t) = 0$ for $t>t_{1}$. Here, $\omega =30$ is the angular frequency of the
incident plane wave.

\textbf{Completing the backscattering data:} We recall that our data are
available only on the backscattering side of the inversion domain, i.e. at $%
\overline{\Omega }\cap \left\{ z=0.04\right\} $. Therefore, the missing
boundary data on the other sides of the rectangular parallelepiped $\Omega $
were approximated by the corresponding simulated data for the homogeneous
medium with $\epsilon \left( \mathbf{x}\right) \equiv 1$ (we recall that
after extracting the targets' signals, as explained in section \ref{sec:extr}%
, the targets were treated as ones placed in air). This approximation has
been used in our previous works and found to provide good reconstruction
results, see \cite{B-K:JIIP2012,BTKF:IP2013,TBKF:SISC2014}.

\subsection{Two tests and stopping criteria}

\label{subsec:twotest}

We have analyzed the performance of the proposed algorithm with two
different tests: \textit{Test 1} and \textit{Test 2}. In Test 1, we made use
of the first tail function as described in section \ref{sec:gc}, with which
the global convergence is rigorously guaranteed. In Test 2, the estimated
burial depth and the $xy$-cross section of the target via the data
preprocessing procedure were used to restrict the domain in which the
coefficient $\epsilon $ was reconstructed and to choose the first tail
function. More precisely, for each target, let $x_{t,min}=\min \{x\in \Gamma
_{T}\},\quad x_{t,max}=\max \{x\in \Gamma _{T}\},$ where $\Gamma _{T}$ is
the estimated $xy$ cross section of the target, see \cite{TBKF:SISC2014} for
how this cross section is estimated. The numbers $y_{t,min}$ and $y_{t,max}$
are defined similarly. Then, we define the extended $xy$ cross section by 
\begin{equation*}
\Gamma _{T,ext}=\{x_{t,min}-0.03<x<x_{t,max}+0.03,\
y_{t,min}-0.03<y<y_{t,max}+0.03\}.
\end{equation*}%
Moreover, denote by $z_{t,front}$ the estimated location of the front side
of the target in the $z$ direction, given by the burial depth estimation. We
then define the following domain $\Omega _{T,ext}$ 
\begin{equation*}
\Omega _{T,ext}:=\{\mathbf{x}\in \Omega :(x,y)\in \Gamma _{T,ext},\
-0.2<z<z_{t,front}+0.02\}.
\end{equation*}%
Clearly, $\Omega _{T,ext}\subset \Omega $. Moreover, this domain should
contain the unknown target we are looking for. The last number $0.02$ was
for compensating for possible error in the estimated burial depth of the
target. Next, we chose the first tail function $V_{0}$ as the function (\ref%
{eq:tail}), where the function $w\left( \mathbf{x},\overline{s}%
\right) $ was computed for the coefficient $\epsilon \left( \mathbf{%
x}\right) :=\epsilon _{0}\left( \mathbf{x}\right) $, where 
\begin{equation*}
\epsilon _{0}(\mathbf{x})=\epsilon _{u},\text{ for }\mathbf{x}\in \Omega
_{T,ext},\ \epsilon _{0}(\mathbf{x})=1,\text{ for }\mathbf{x}\notin \Omega
_{T,ext}.
\end{equation*}%
In this paper, the upper bound $\epsilon _{u}$ for the function $\epsilon
\left( \mathbf{x}\right) $ was chosen as $\epsilon _{u}=25$.

Although the convergence of the resulting algorithm for Test 2 has not been
rigorously proved yet, our numerical results show good reconstructions, see
also \cite{TBKF:SISC2014} for results when targets are in air. Note that we
did not use a priori information about the targets. Instead, the information
used in choosing the first tail function was derived from data preprocessing.

\textbf{Stopping criteria}: As mentioned in Remark \ref{re:1}, stopping
criteria of the algorithm should be addressed numerically. In this paper, we
used the stopping criteria proposed in \cite{BTKF:IP2013,TBKF:SISC2014}. We
briefly recall these criteria here for the reader's convenience.

\textit{Stopping criterion of Test 1:} The inner iteration with respect to $i
$ is stopped at $i=m_{n}$ such that 
\begin{equation*}
\text{{}}D{_{n,i}}\geq D{_{n,i-1},}\text{ or }i<i_{max},
\end{equation*}%
where $D_{n,i}=||{V}_{n,i}|_{\Gamma _{p}}-{V}_{prop}||_{L_{2}(\Gamma _{p})}.$
Here $\Gamma _{p}$ is the backscattering side of $\Omega $ and $V_{prop}$ is
the tail function computed from the propagated data at $\Gamma _{p}$, and $%
i_{max}$ is the maximum number of inner iterations. In Test 1, we have
chosen $i_{max}=8$.

The outer iteration with respect to the pseudo  frequency is stopped when
the error function $D_{n,1}$ attains the first local minimum with respect to 
$n$.

\textit{Stopping criterion of Test 2:} The inner iteration is stopped using
the same criterion as in Test 1 but with $i_{max} = 5$. The outer iteration
is stopped when the error function $D_{n,m_n}$, i.e., the error function at
the final inner iteration, attains the first local minimum.

We have observed in our tests that these stopping criteria gave good results
for non blind targets.

\subsection{Summary of reconstruction results and discussion}

In Tables \ref{ta:2} and \ref{ta:3} we summarize reconstruction results of
the two tests for the data sets listed in Table \ref{ta:1}. Table \ref{ta:2}
shows the results for the non metallic targets. For these targets, the
refractive index $n($target$)=\sqrt{\epsilon (\text{target})}$ ($\epsilon (%
\text{target})$ was chosen by $\epsilon (\text{target})=\max_{\mathbf{x}\in
\Omega }\epsilon (\mathbf{x})$) is shown instead of the dielectric constant $%
\epsilon $ because $n($target$)$ was directly measured after
computations were performed. Table \ref{ta:3} shows the burial depths and
the effective dielectric constants of the metallic targets.

\begin{table}[tbp]
\caption{{\protect\small \emph{Result of the globally convergent algorithm:
the refractive indices $n = \protect\sqrt{\protect\epsilon}$ and the burial
depths of non-metallic targets. Object \#11 is a heterogeneous target with
diffuse scattering, see below. Object \#12 is a heterogeneous one with outer
and inner layers, the computed $n$ is compared with the average measured $%
n=1.28$. Object \#23 consists of two targets: wet wood and empty bottle
filled with air. ``Comp.'' stands for ``Computed''. The average error of
strong targets is 8.5\% for Test 1 and 14.7\% for Test 2. The average error
of weak targets is 21.6\% for Test 1 and 13\% for Test 2. }}}
\label{ta:2}
\begin{center}
{\footnotesize \ 
\begin{tabular}{|c|l|r|r|l|l|l|}
\hline
Object & Material & Comp. & Exact & Comp. & Comp. & Measured \\ 
\# &  & depth & depth  & $n$, Test 1 & $n$, Test 2 & $n$ \\ \hline
3 & Water & 3.6 cm & 4.0 cm& 4.7 & 4.9 & 4.88 \\ \hline
4 & Wet wood & 5.5 cm & 9.8 cm & 4.4 & 4.5 & 4.02 \\ \hline
8 & Air & 2.8 cm & 3.0 cm & 1.0 & 0.98 & 1.0 \\ \hline
9 & Teflon & 2.9 cm & 2.5 cm & 1.0 & 1.18 & 1.0 \\ \hline
10 & Ceramic & 4.0 cm & 5.0 cm & 1.0 & 1.23 & 1.39 \\ \hline
11 & Wood with & 4.6 cm & 4.0  cm& 1.0 & 1.46 & 1.89 (wood) \\ 
& metal screws &  &  &  &  & N/A: diffuse scattering \\ \hline
12 & Geode & 2.1 cm & 2.5 cm & 1.0 & 1.52 & 1.31 (outer) \\ 
& (two layers) &  &  &  &  & 1.25 (inner) \\ 
&  &  &  &  &  & 1.28 (average) \\ \hline
13 & Rock & 2.0 cm & 2.3  cm& 1.0 & 1.34 & 1.34 \\ \hline
14 & Coffee grounds & 2.0  cm & 2.5 cm & 1.0 & 1.46 & 1.11 \\ \hline
15 & Ceramic & 2.6 cm & 2.5 cm & 1.0 & 1.51 & 1.39 \\ \hline
19 & Water & 7.5 cm & 9.5 cm & 4.5 & 5.2 & 4.88 \\ \hline
22 & Wet wood & 2.9 cm & 3.0 cm & 4.8 & 5.3 & 4.02 \\ \hline
23 & Wet wood & 5.7  cm& 7.5 cm& 4.0 & 4.1 & 4.02 \\ 
& Empty bottle (air) & missed & 7.5 cm & missed & missed & 1.0 \\ \hline
24 & Wet wood & 5.1 cm & 6.8 cm & 3.67 & 3.0 & 4.02 \\ \hline
\end{tabular}
}
\end{center}
\end{table}

\begin{table}[tbp]
\caption{{\protect\small \emph{Result of the globally convergent algorithm:
the estimated effective dielectric constants and the burial depths of
metallic targets. Object \#18 is a heterogeneous one: a wooden dool with a
metallic block inside. Object \#20 consists of two targets: a metallic block
and a rock. Measured $n\left( \text{rock}\right) =1.34.$ Objects \#5, 16, 17
consist of two metallic targets. Object \#25 consists of two metallic
targets with 1 cm distance between their surfaces: super resolution, see
Figure 5.4.}}}
\label{ta:3}
\begin{center}
{\footnotesize \ 
\begin{tabular}{|c|l|r|r|l|l|}
\hline
Object & Material & Computed & Exact & Computed $\epsilon$ & Computed $%
\epsilon$ \\ 
\# &  & depth  & depth  & Test 1 & Test 2 \\ \hline
1 & Metal & 2.9 cm & 4.0 cm & 29.9 & 46.4 \\ \hline
2 & Metal & 2.9 cm & 3.0 cm & 24.5 & 31.0 \\ \hline
5 & Metal & 3.0 cm & 3.0 cm & 23.4 & 32.4 \\ 
& Metal & 3.6 cm & 3.0 cm & 30.5 & 41.2 \\ \hline
6 & Metal & 2.8 cm & 8.5 cm & 27.8 & 37.5 \\ 
& Teflon & missed & 8.5 cm &  &  \\ \hline
7 & Metal & 9.9 cm & 14.0 cm & 47.4 & 65.8 \\ \hline
16 & Metal & 2.5 cm & 4.5 cm & 19.9 & 24 \\ 
& Metal & 3.7 cm & 4.5 cm & 33.7 & 47.5 \\ \hline
17 & Metal & 2.0 cm & 3.8 cm & 30.0 & 51.1 \\ 
& Metal & 2.7 cm & 3.8 cm & 54.8 & 93.5 \\ \hline
18 & Wood, metal block inside & 7.1 cm & 8.5 cm & 18.3 & 19.9 \\ \hline
20 & Metal & 6.8 cm & 8.5 cm & 30.0 & 48.1 \\ 
& Rock & missed & 8.5 cm &  &  \\ \hline
21 & Metal & 5.1 cm & 7.5 cm & 22.1 & 28.2 \\ \hline
25 & Metal & 3.8 cm & 4.0 cm & 70.0 & 99.8 \\ 
& Metal & 4.0 cm & 4.0 cm & 40.8 & 56.5 \\ \hline
\end{tabular}
}
\end{center}
\end{table}

As described in section \ref{sec:ext}, the burial depth was estimated based
on the time delay between the reflection by the sand's surface and the
target's signal. Note that our incident signal was not really a short
impulse. It is therefore natural to expect some level of error in our
estimates. Since we made use of peaks of the signals in estimating the
depth, the error we expect is about the distance between two consecutive
peaks, which is equal to half of the wavelength (2 cm).

%
%
%

From Tables \ref{ta:2} and \ref{ta:3} we can see that the burial depth was
accurately estimated in most cases, with the errors not exceeding 2 cm.
There are two cases (\#4, 7) in which the errors were about 4 cm. These
targets were buried at rather deep depths of about the limiting depth (10
cm) for antipersonnel land mines. This made the estimate less accurate
because of possible uncertainty in measuring the refractive index of the
sand. Also, there might be an error in recording the exact burial depths
during the data acquisition for deeply buried targets.

The estimates of the refractive indices of non-metallic targets with
refractive indices larger than that of the sand (water and wet wood) are
quite accurate with the average error of about 9.7\% for Test 1 and 15.2\%
for Test 2. Note that the error in our direct measurement of the refractive
index of the wet wood was 10\%. For water, we were unable to directly
measure its refractive index at the used quite high frequency of the signal,
which was about 7.5 GHz. Therefore, we have made a separate experiment: we
have placed that bottle of water in air, measured the backscattering data
and then reconstructed the refractive index using the globally convergent
method as in \cite{BTKF:IP2013, TBKF:SISC2014}. The result of $n=4.88$
matches well the experimentally measured refractive index of 4.84 at high
frequencies in Table 3.1 of \cite{Farr}. Moreover, by comparing our computed 
$n$ for water in Table \ref{ta:2} with this reference value $n=4.88$, we can
see the consistency of our results.

Targets with smaller refractive indices than that of the sand are of
interest since they are models of plastic land mines and IEDs. We have
observed that we can image these targets only if their burial depths do not
exceed 5 cm. The average error shown in Table \ref{ta:2} for these weak
targets is about 21.6\% for Test 1 and 13\% for Test 2. The average measurement
error of $n$ for weak targets was about 5.4\%.

In our experiments, we have missed some weak targets (not shown here), which
had more than 5 cm burial depths. For these weak targets, we have observed
that their signals were blended by the reflection from the sand's surface.
Therefore, we could not detect any target's signal out of them. Note that,
since our current algorithm uses the Laplace transform, it is applicable
only when we can detect and extract targets' signals and remove the
reflection by the sand's surface as well as noise at earlier times.
Otherwise, they will dominate the targets' signals after the Laplace
transform, see Figure \ref{fig:4.4}. Thus, these missed cases were not due
to the inversion algorithm.

The signals of the metallic targets were strong compared to the sand's
signal. Therefore they were quite easy to detect. We recall that metallic
targets can be approximated by dielectric ones with large \textit{effective}
dielectric constants. The estimated effective dielectric constants of our
metallic test targets are between 20 and 100. In our previous works, we have
established that the effective dielectric constant of metals should be
larger than 10--15, see \cite{BTKF:IP2013,K-B-K-S-N-F:IP2012, TBKF:SISC2014}%
. Therefore our results here are compatible with our previous studies.

\begin{figure}[tph]
\centering
\subfloat[Test 1, 3D view]{\includegraphics[width=0.4\textwidth,
height=0.37\textwidth]{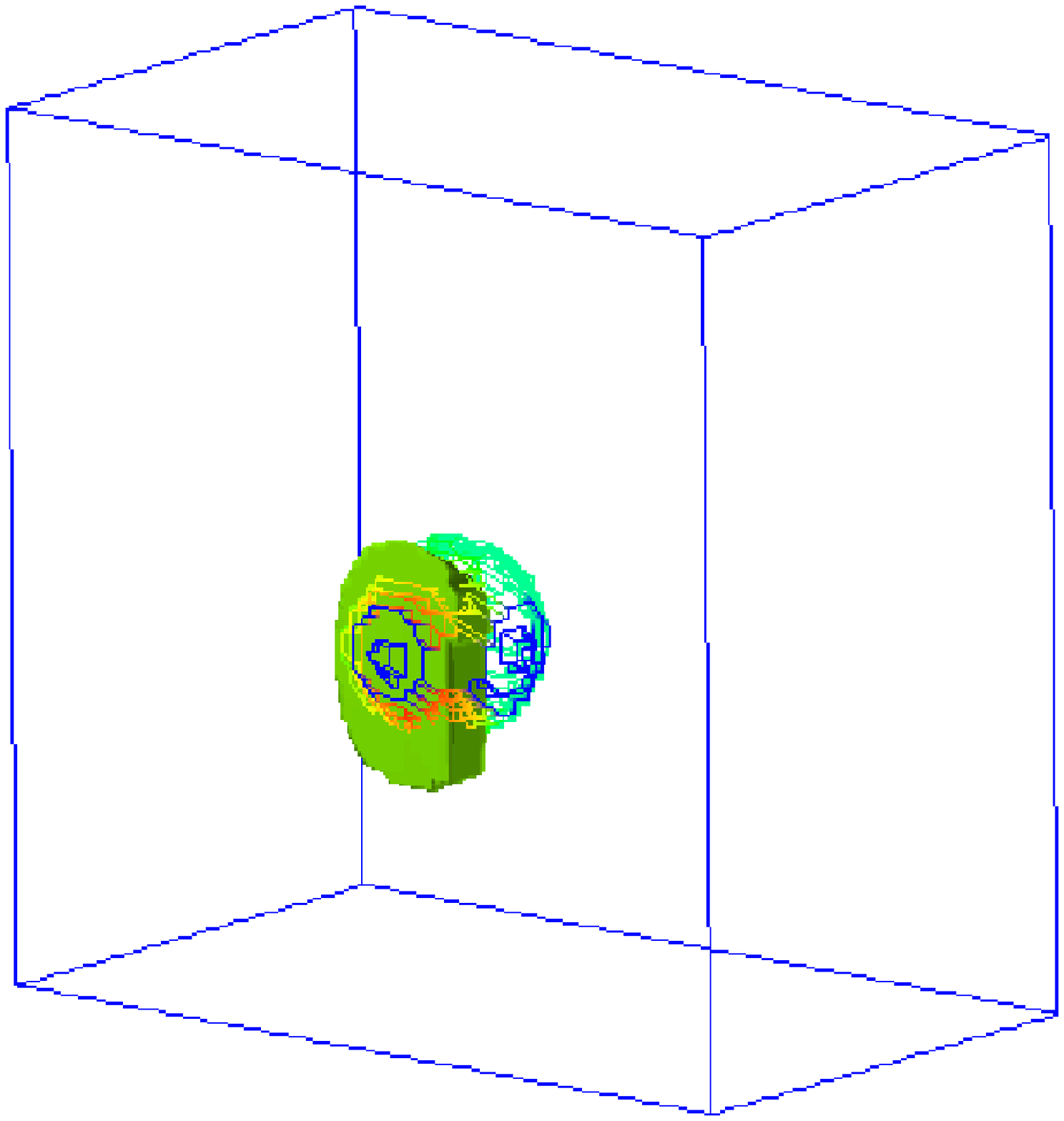}} 
\subfloat[Test 2, 3D
view]{\includegraphics[width=0.4\textwidth,
height=0.37\textwidth]{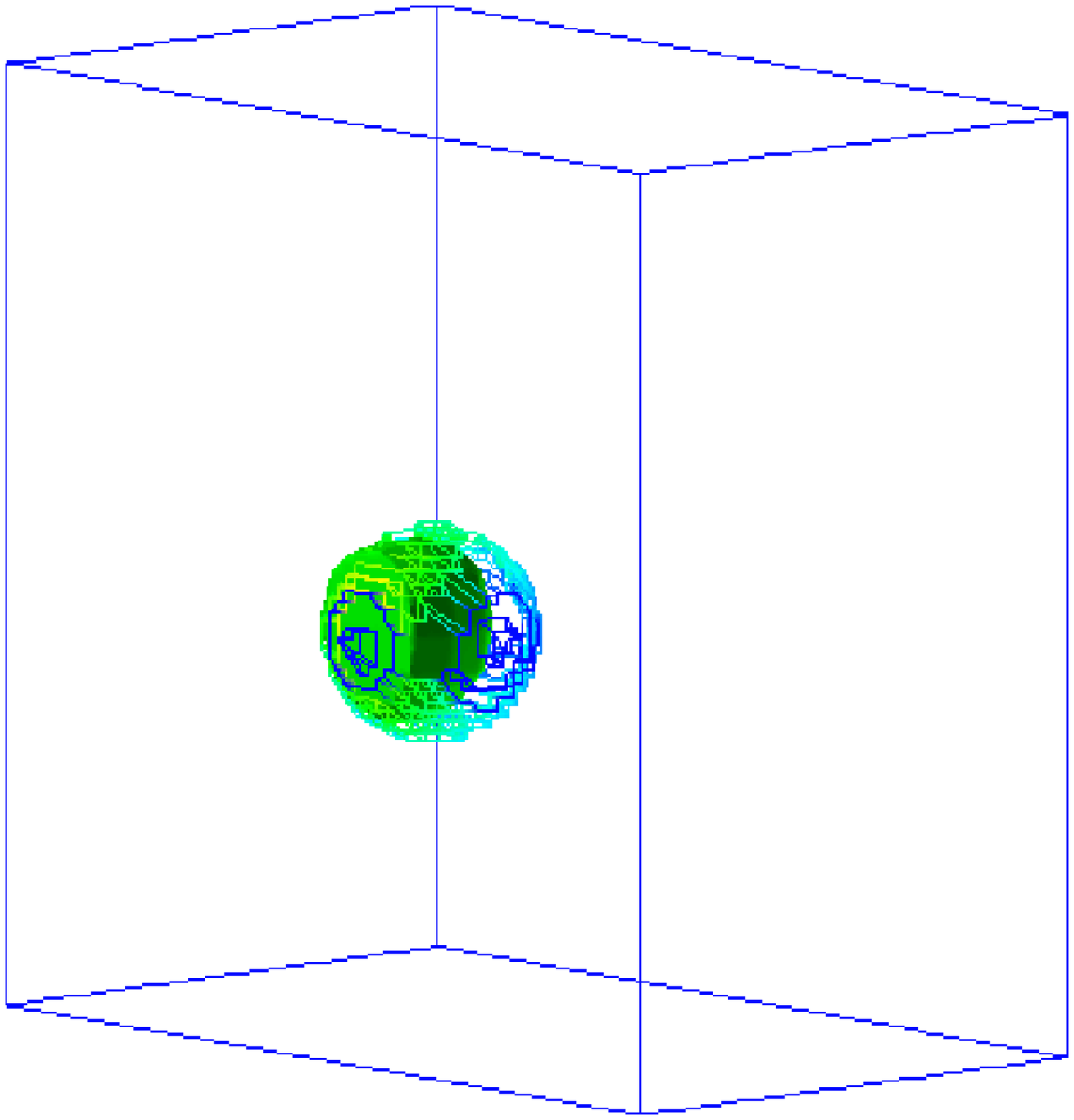}}
\par
\subfloat[Test 1, $xy$
view]{\includegraphics[width=0.4\textwidth, height=0.37\textwidth]{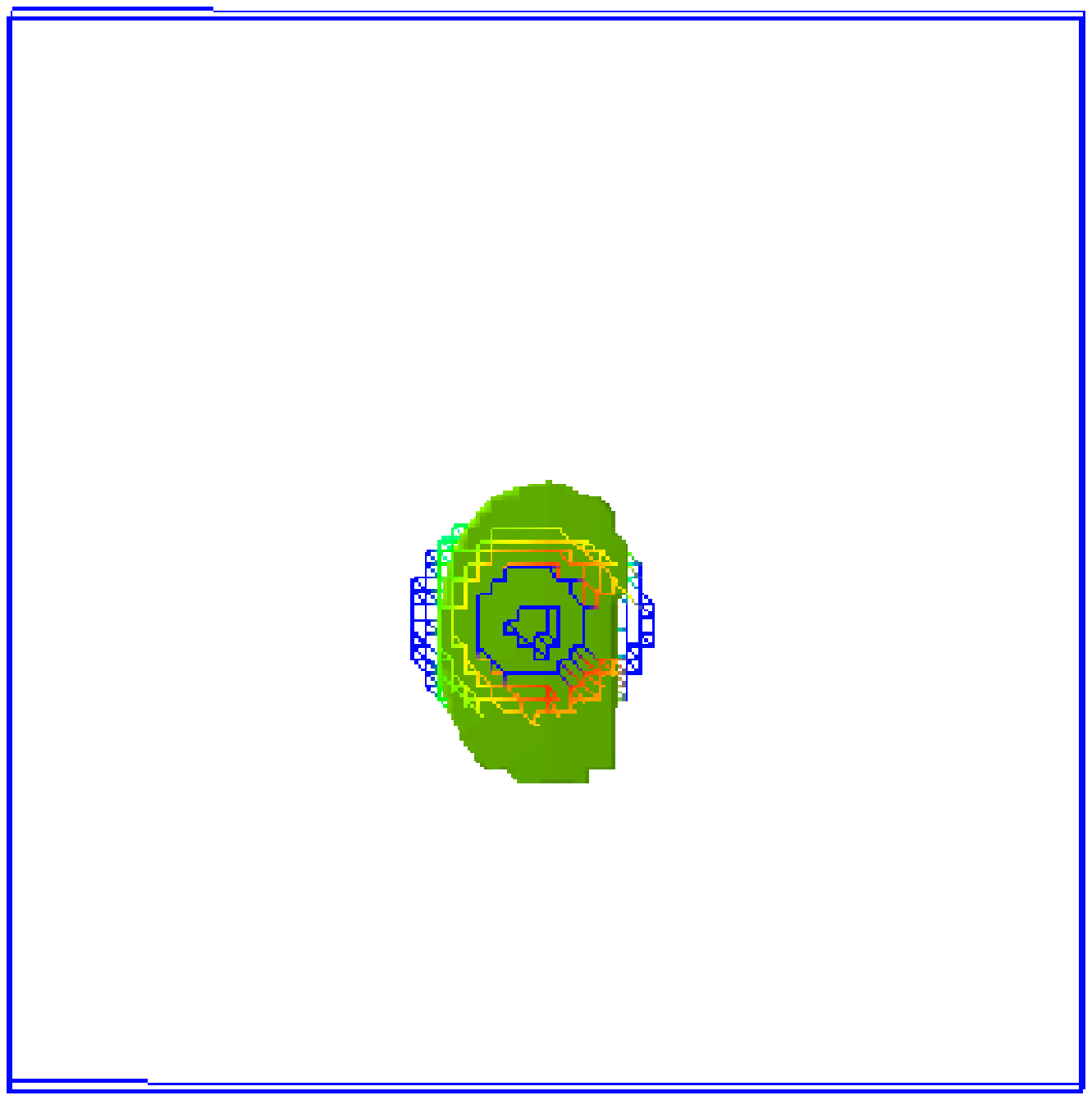}}
\subfloat[Test 2, $xy$
view]{\includegraphics[width=0.4\textwidth, height=0.37\textwidth]{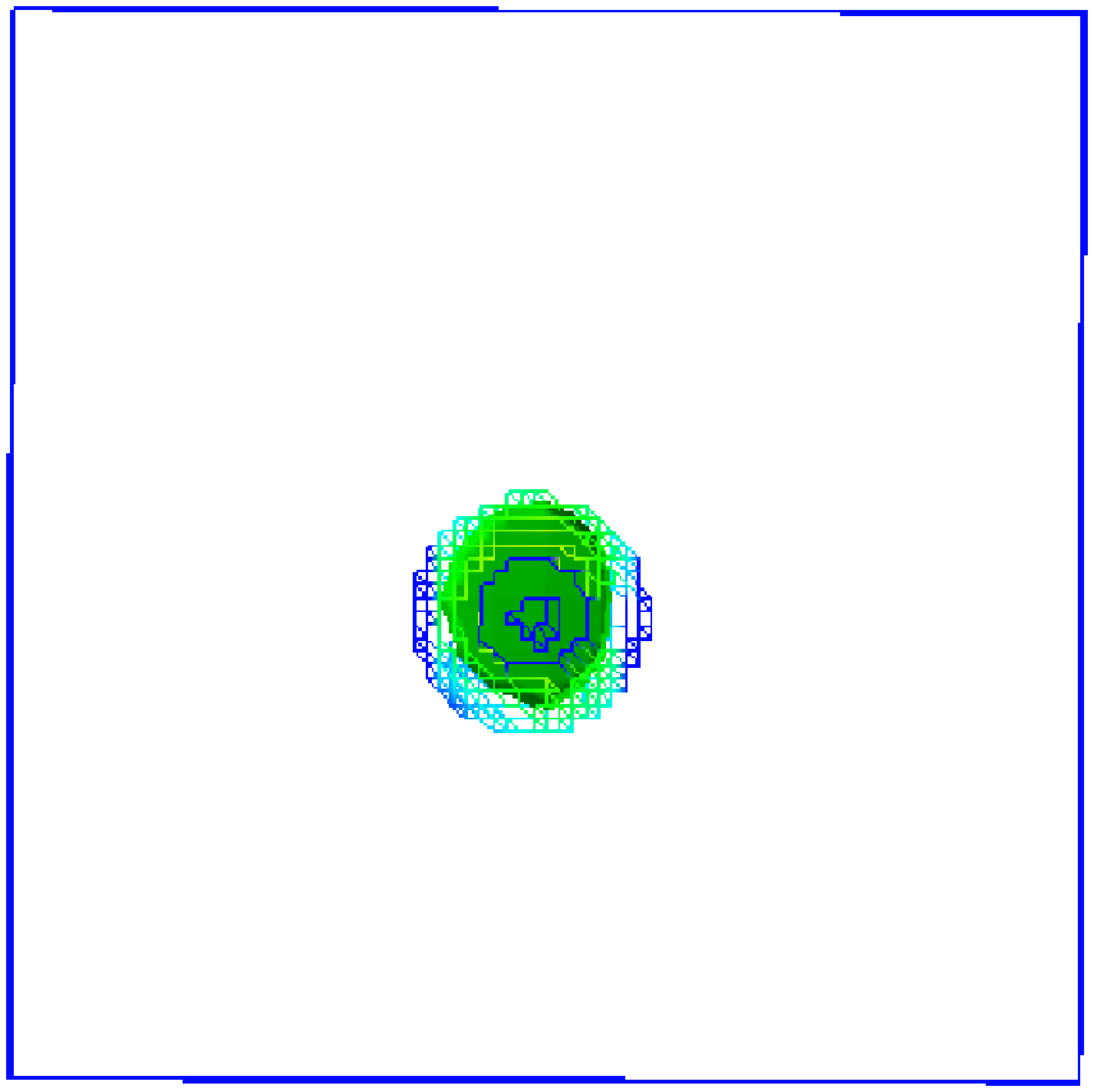}}
\caption{Reconstructed shape of Target \#2 (a metallic ball): $xy$ view
means the projections of the target on the $xy$ plane. The thin lines
indicate the true shape.}
\label{fig:shape0}
\end{figure}

\begin{figure}[tph]
\centering
\subfloat[Test 1, 3D view]{\includegraphics[width =
0.4\textwidth, height = 0.37\textwidth]{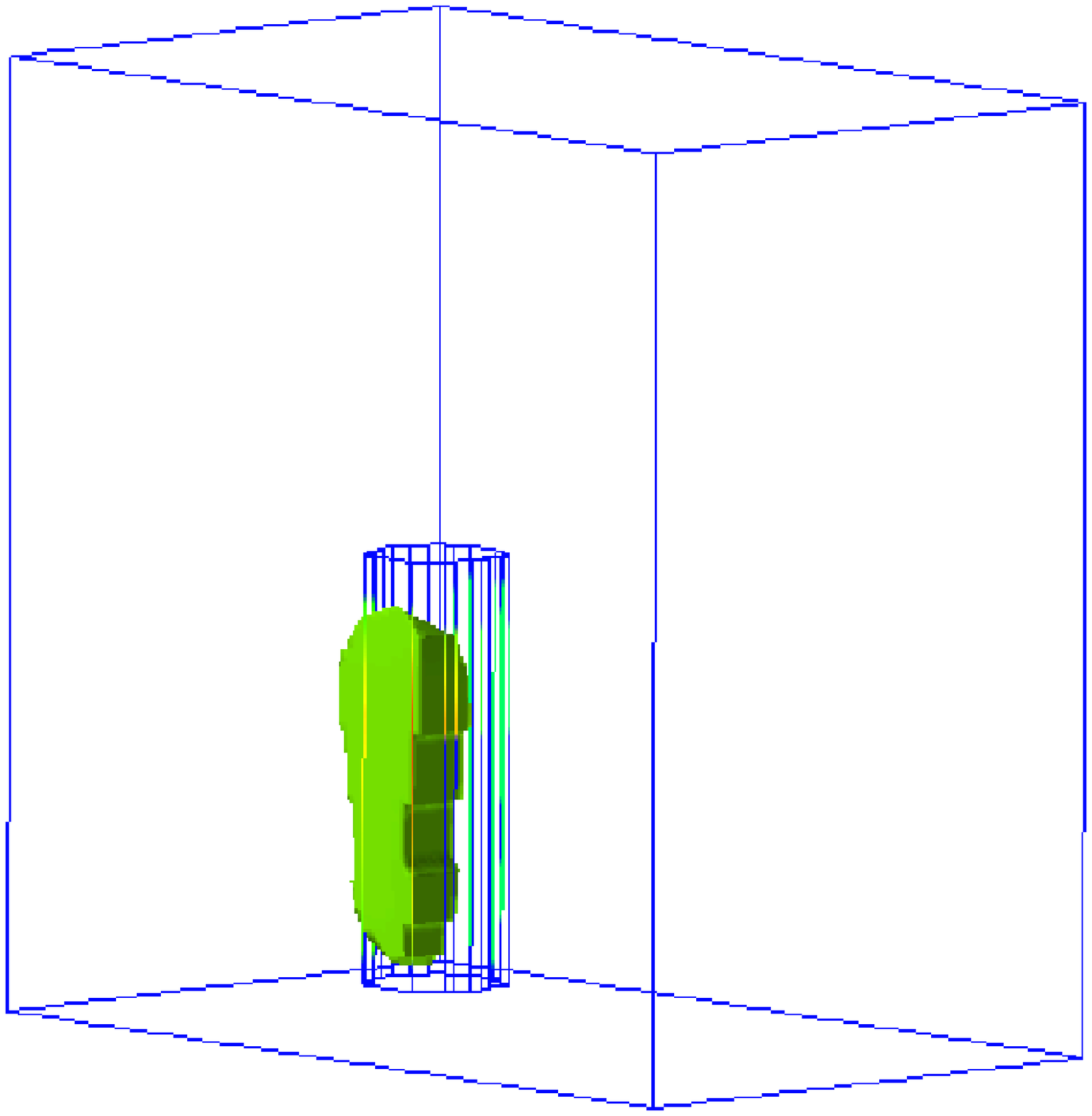}} 
\subfloat[Test 2, 3D view]{\includegraphics[width =
0.4\textwidth, height = 0.37\textwidth]{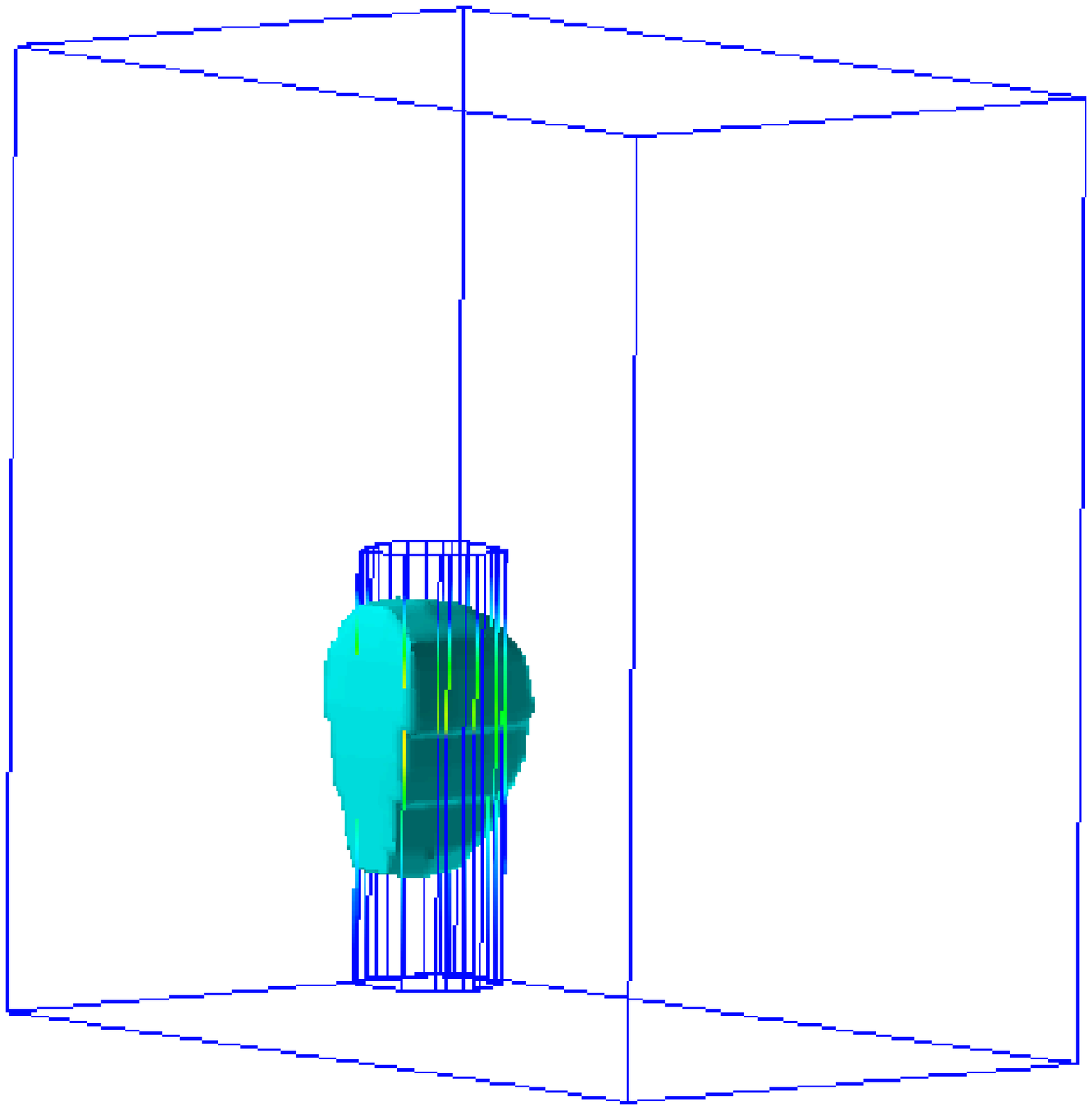}}
\par
\subfloat[Test 1, $xy$
view]{\includegraphics[width =
0.4\textwidth, height = 0.37\textwidth]{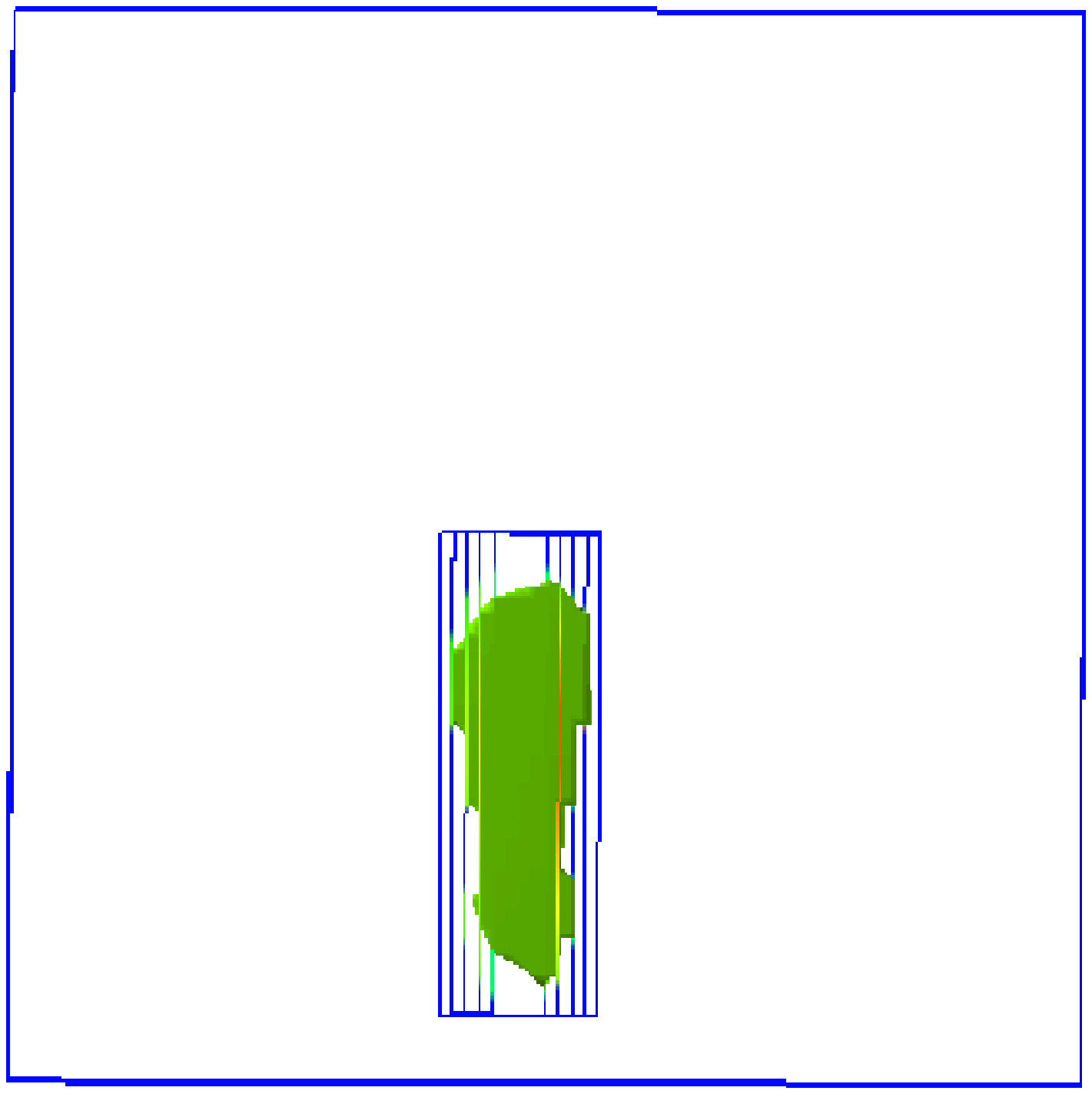}} 
\subfloat[Test 2, $xy$
view]{\includegraphics[width =
0.4\textwidth, height = 0.37\textwidth]{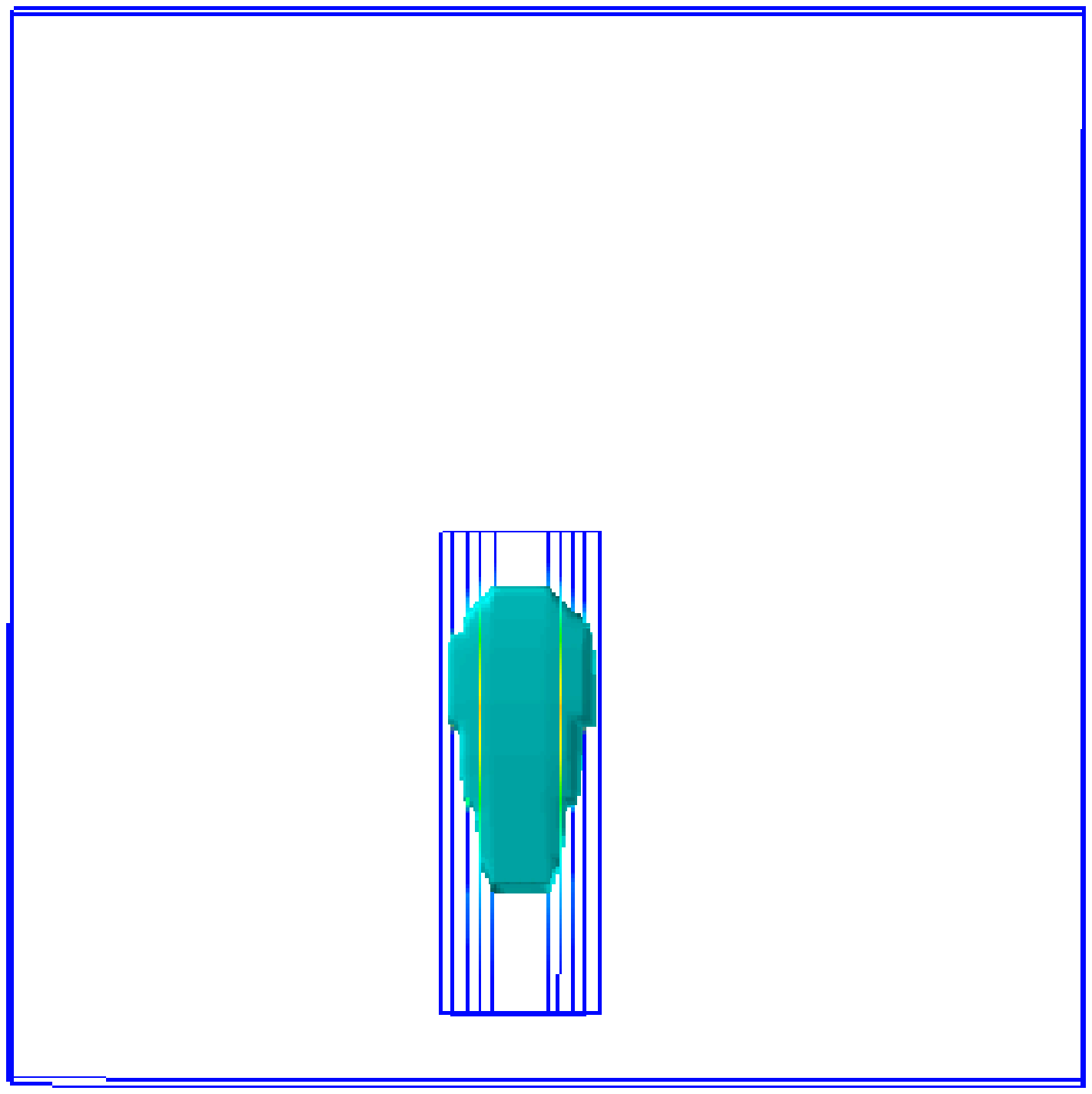}} 
\caption{Reconstructed shape of Target \#3 (a bottle of water). The thin
lines indicate the true shape.}
\label{fig:shape1}
\end{figure}

\begin{figure}[tph]
\centering
\subfloat[Test 1, 3D view]{\includegraphics[width =
0.4\textwidth, height = 0.37\textwidth]{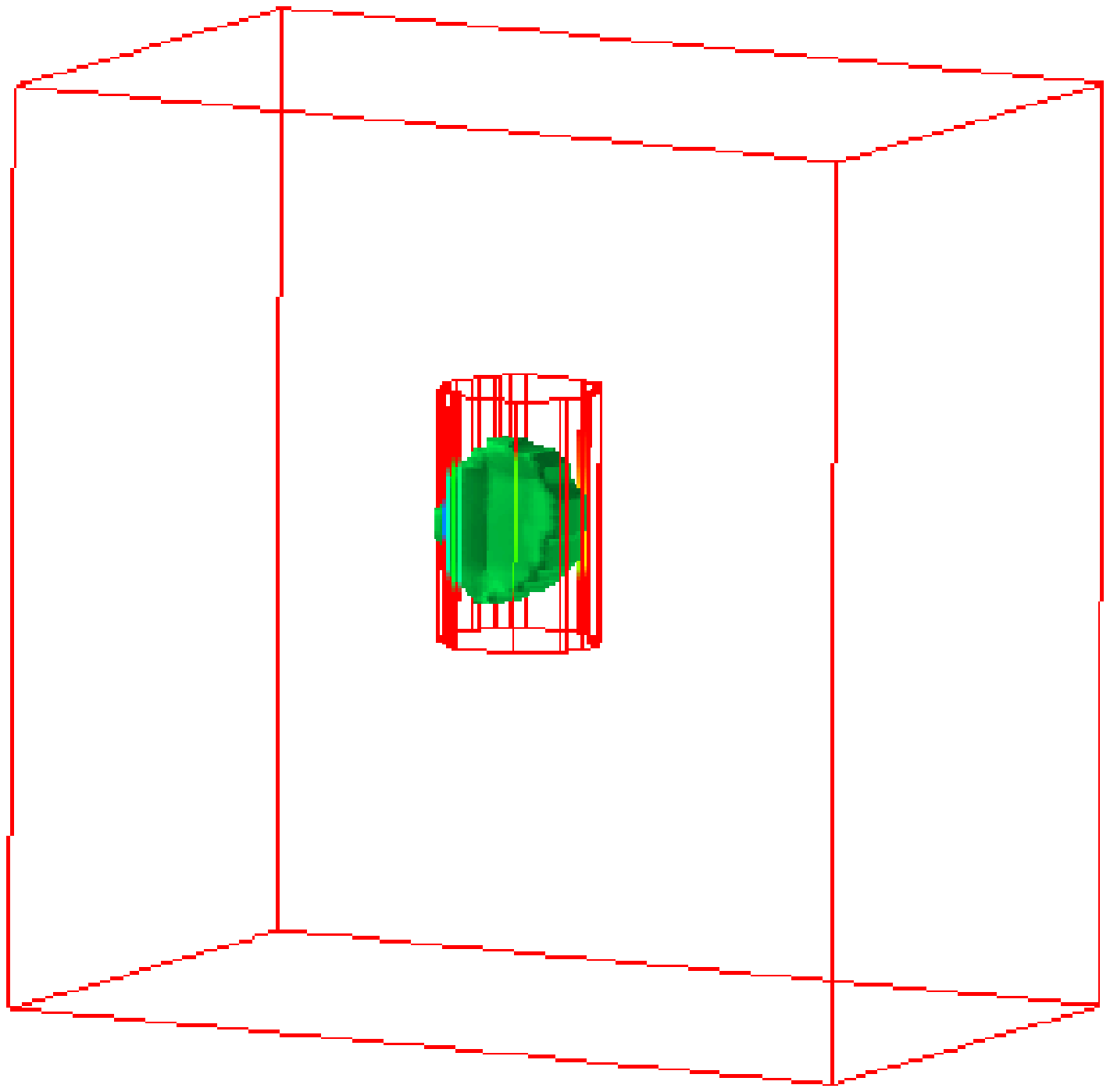}} 
\subfloat[Test 2, 3D view]{\includegraphics[width =
0.4\textwidth, height = 0.37\textwidth]{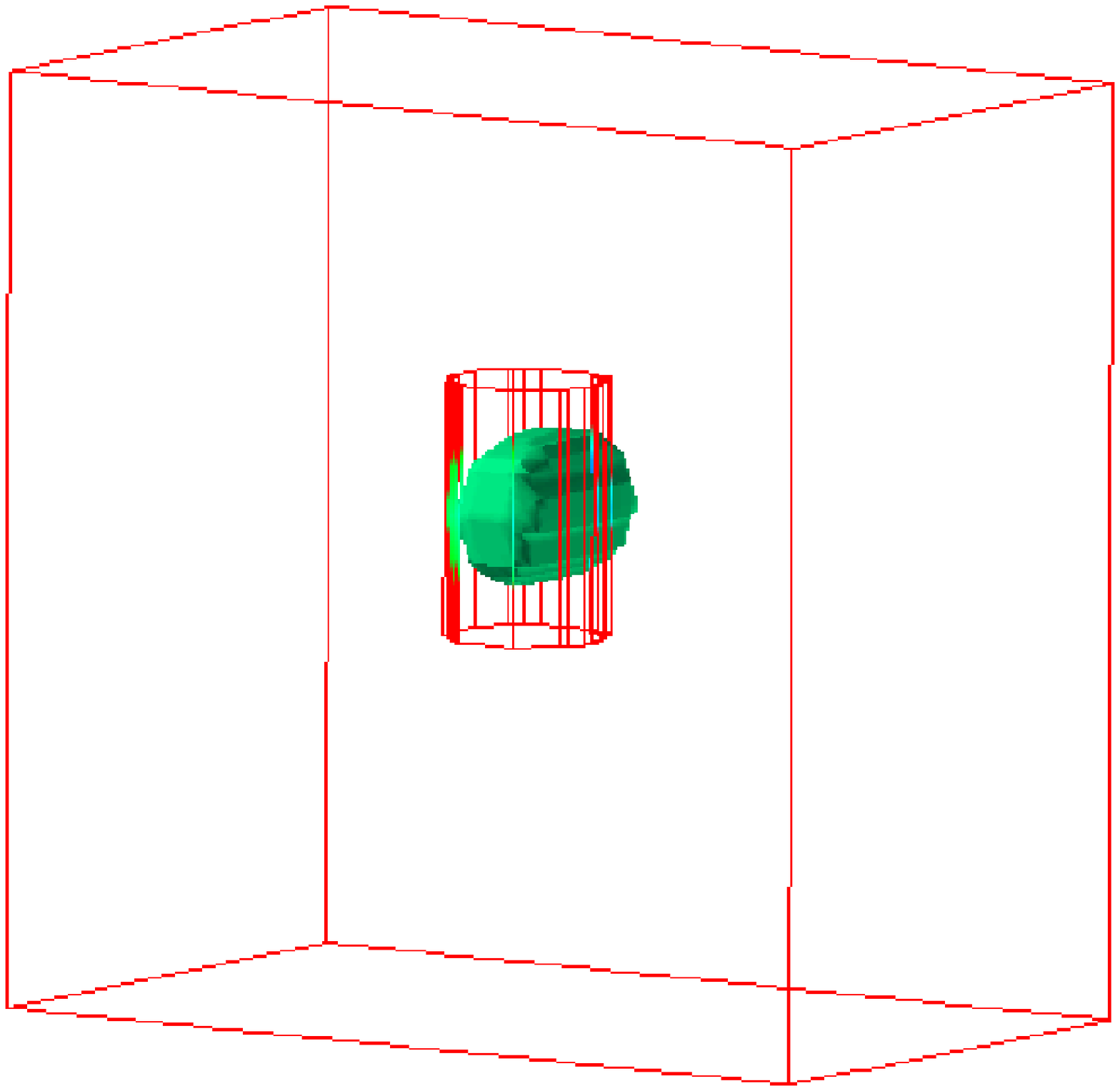}}
\par
\subfloat[Test 1, $xy$
view]{\includegraphics[width =
0.4\textwidth, height = 0.37\textwidth]{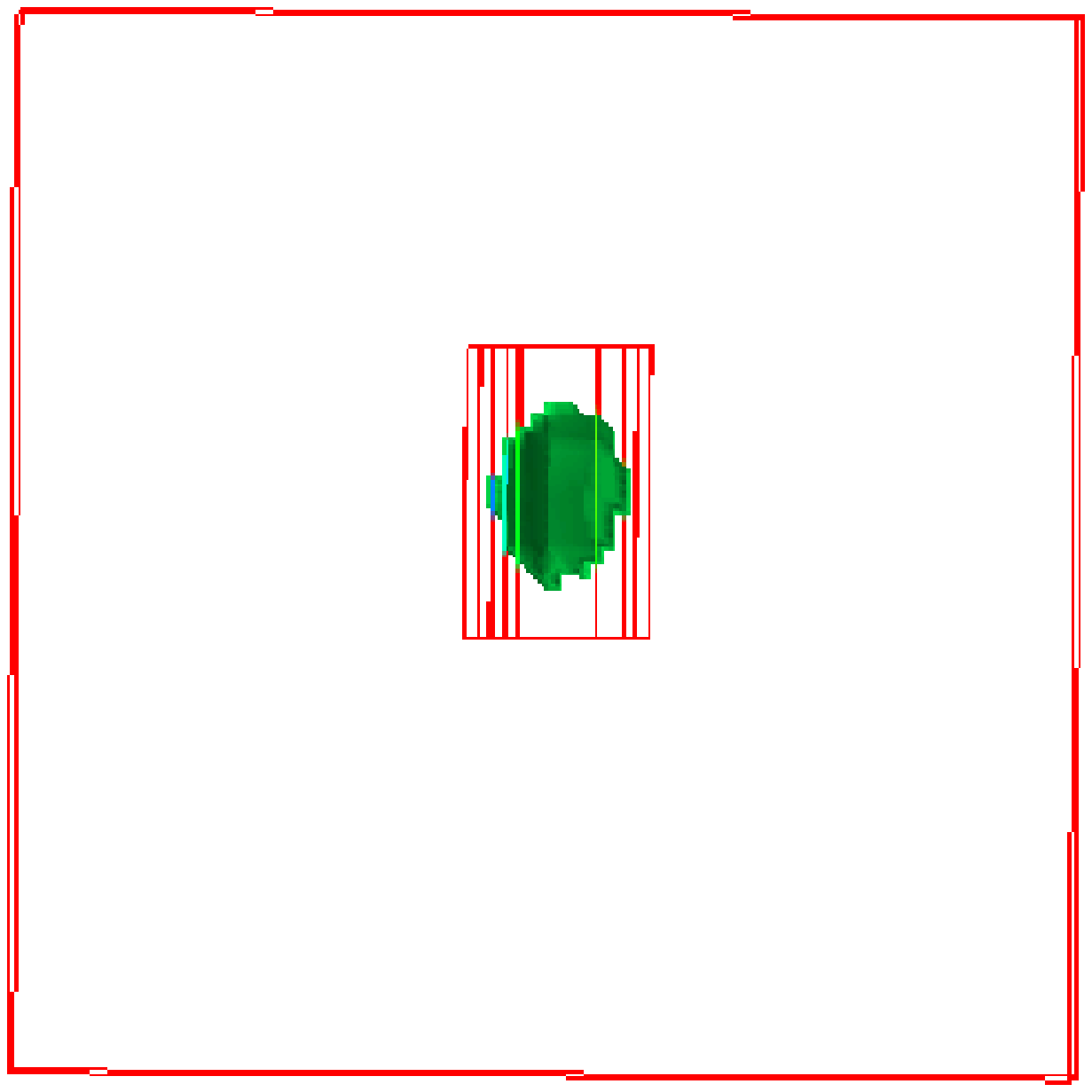}} 
\subfloat[Test 2, $xy$
view]{\includegraphics[width =
0.4\textwidth, height = 0.37\textwidth]{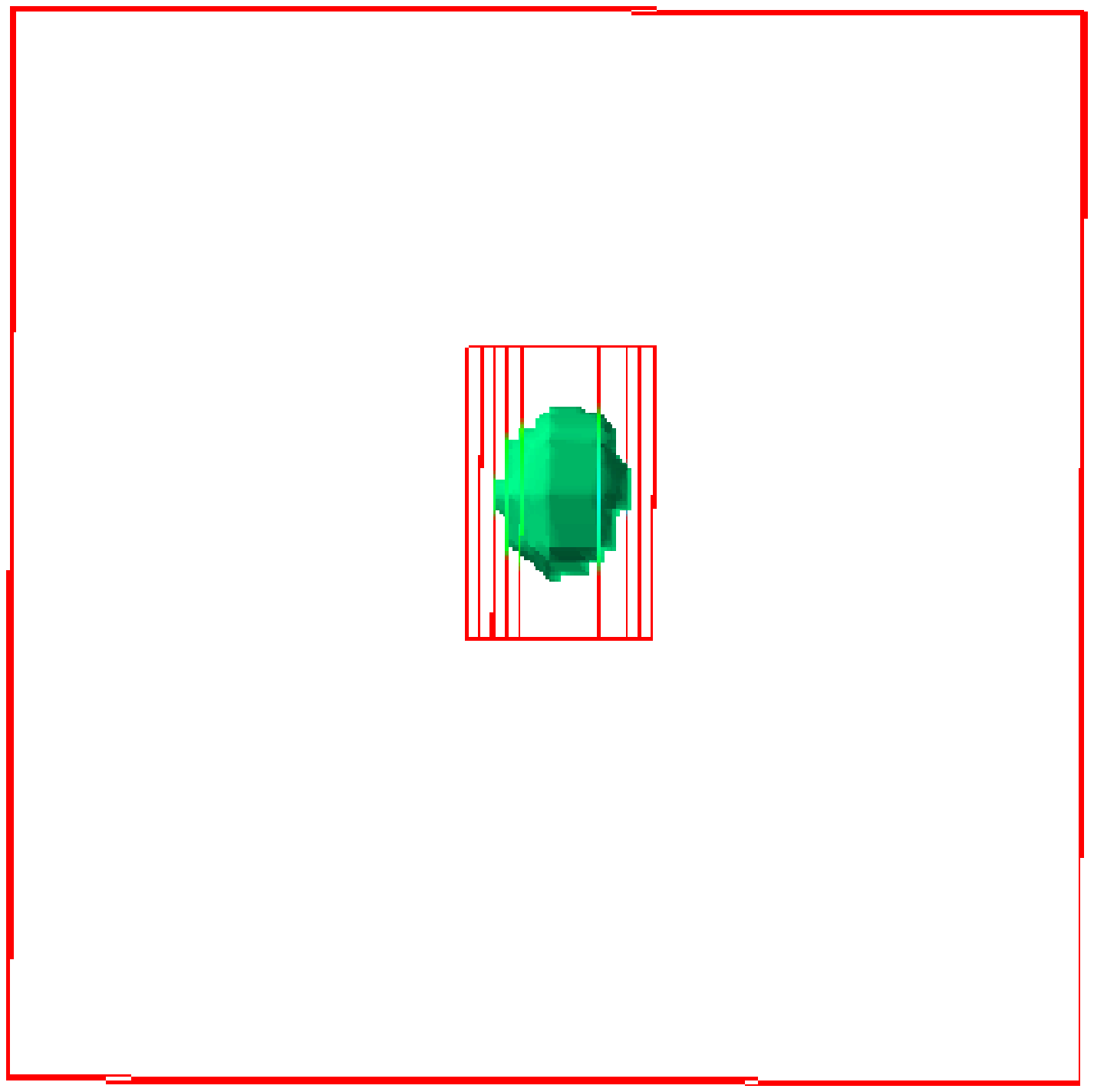}} 
\caption{Reconstructed shape of Target \#10 (a ceramic mug). The thin lines
indicate the true shape.}
\label{fig:shape2}
\end{figure}

\begin{figure}[tph]
\centering
\subfloat[Test 1, 3D view]{\includegraphics[width=0.4\textwidth, height =
0.37\textwidth]{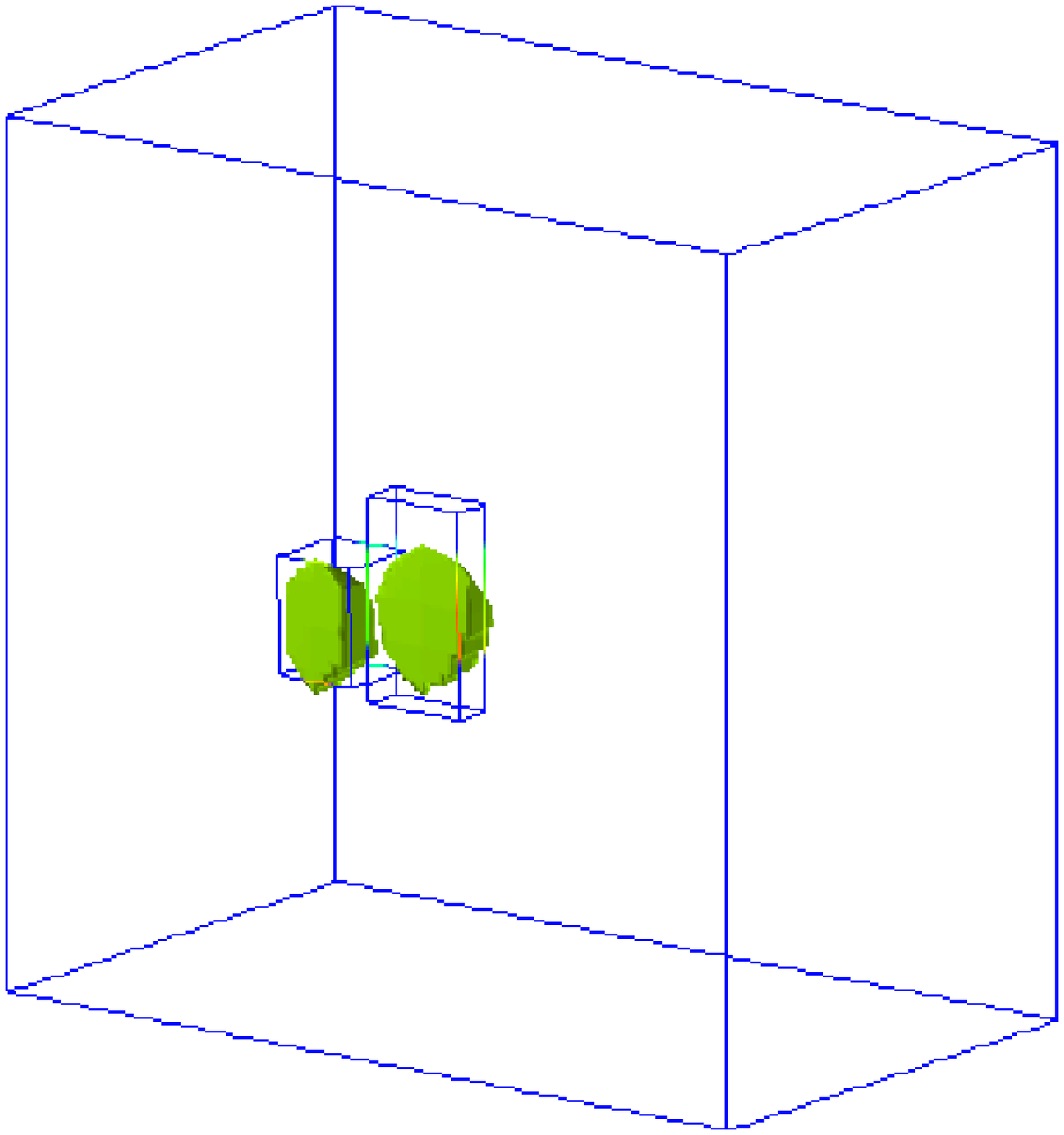}} 
\subfloat[Test 2, 3D
view]{\includegraphics[width =0.4\textwidth, height =
0.37\textwidth]{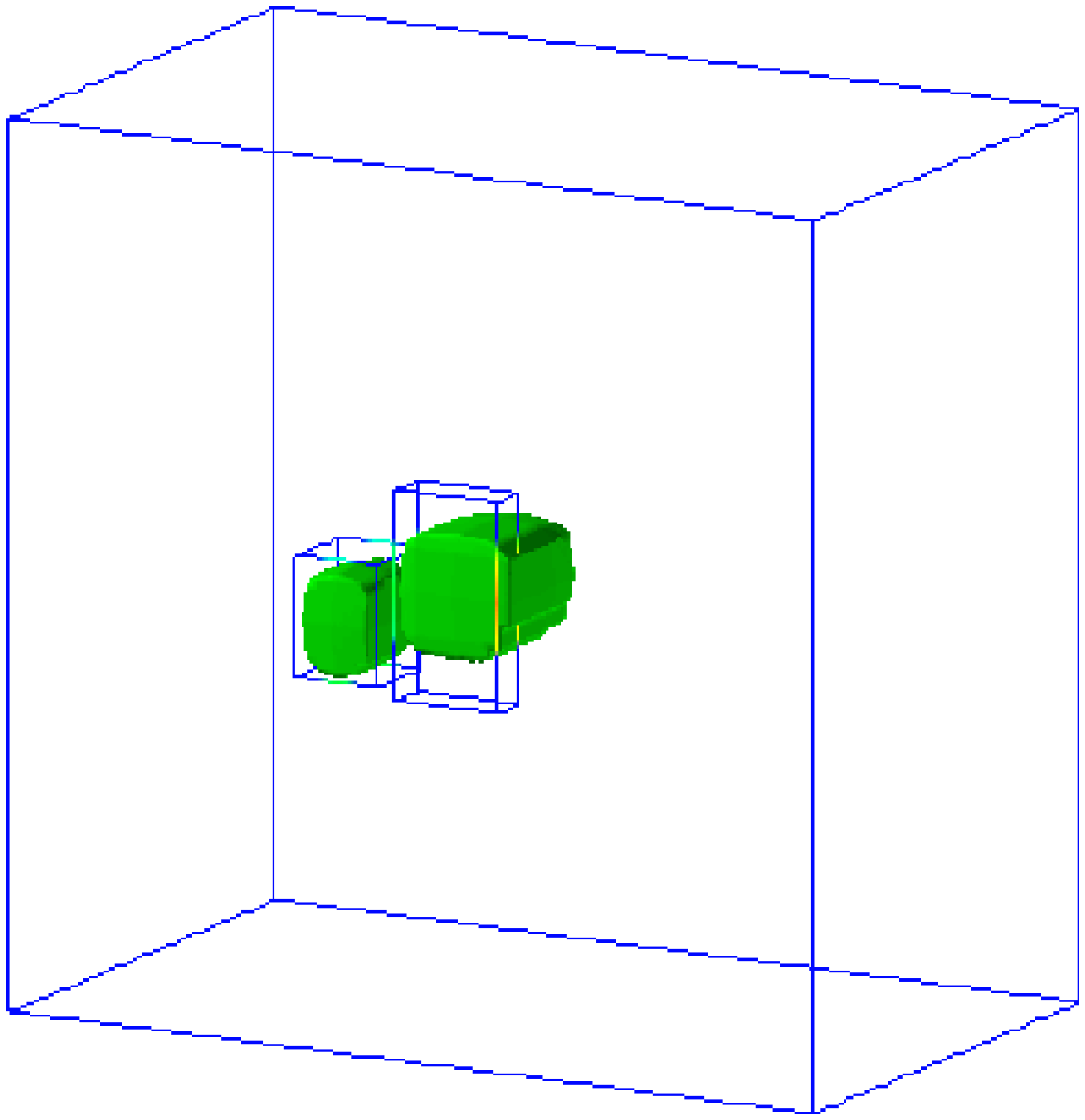}}
\par
\subfloat[Test 1, $xy$ view]{\includegraphics[width =0.4\textwidth, height
= 0.37\textwidth]{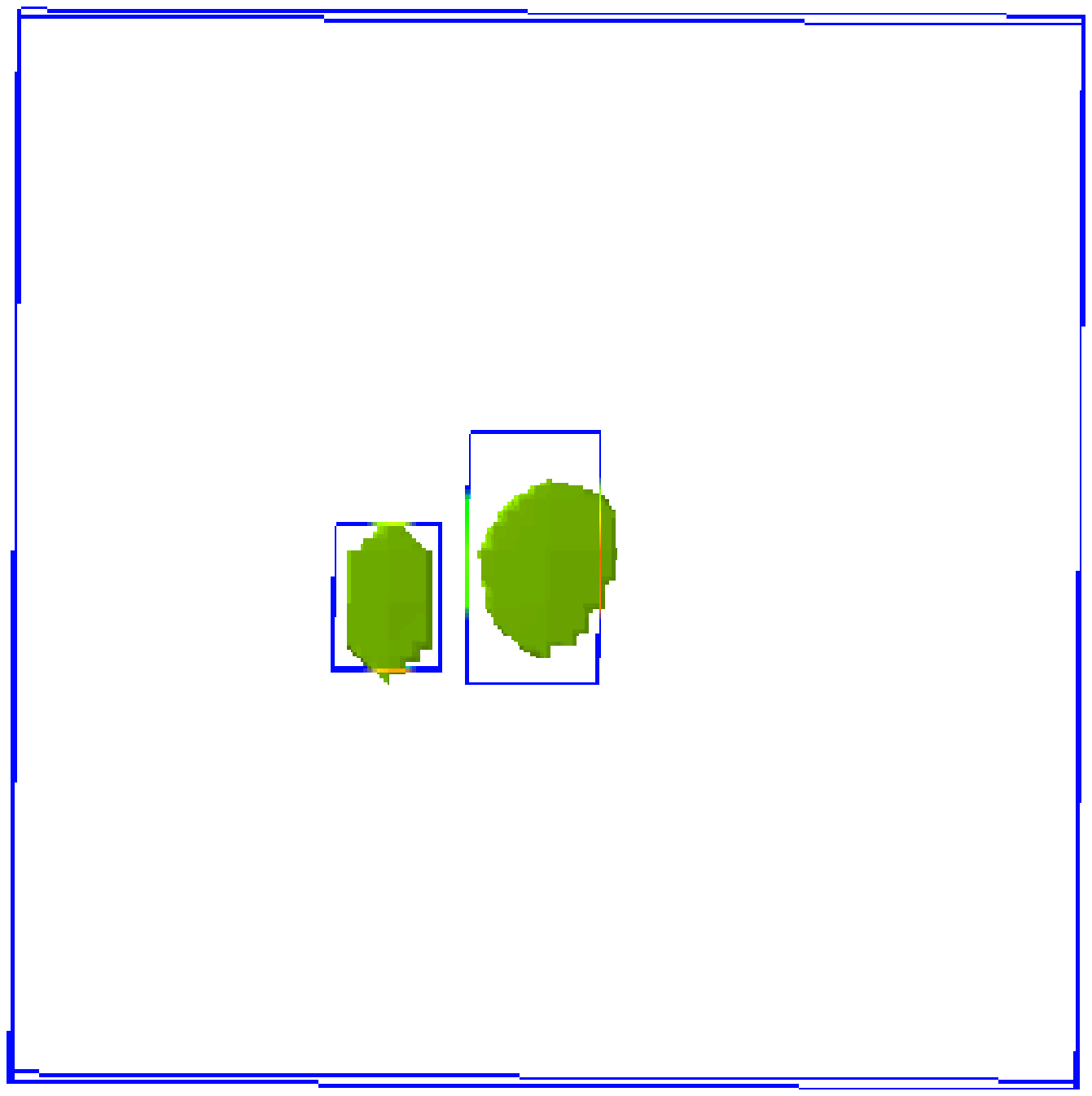}} 
\subfloat[Test 2, $xy$
view]{\includegraphics[width =0.4\textwidth, height =
0.37\textwidth]{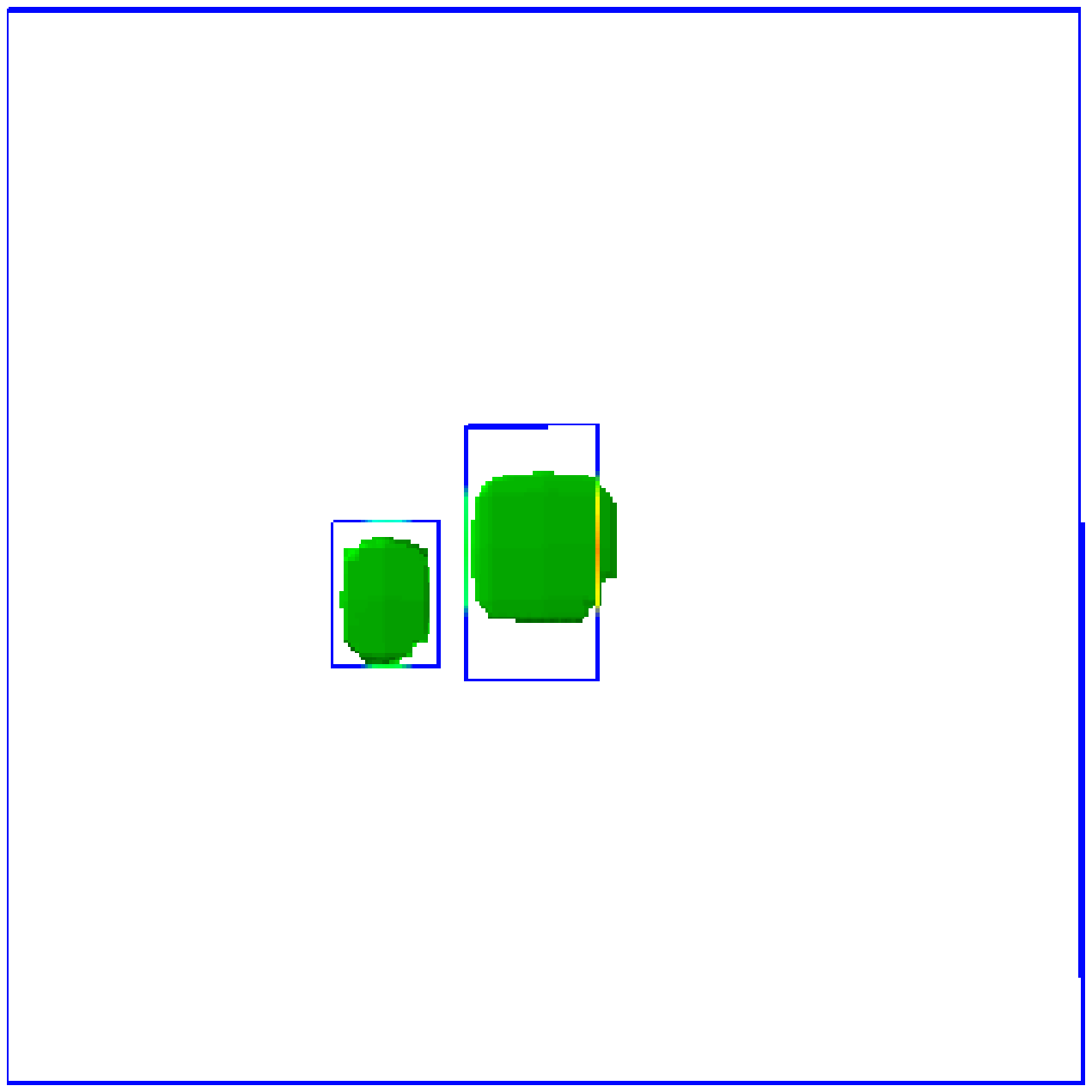}} 
\caption{Reconstructed shape of Target \#25 (two metallic blocks at 1 cm
separation). The thin lines indicate the true shape.}
\label{fig:shape3}
\end{figure}

In our tests, there were four cases (\#5, 16, 17, 25 in Table \ref{ta:3}) in
which there were two metallic targets simultaneously. In each case we have
accurately imaged both targets. In particular, in the data set \#25, the two
metallic blocks were at 1 cm distance, see Figure \ref{fig:shape3}. On the
other hand, the wavelength of our device is about 4 cm. Thus, the \textit{%
super resolution} is achieved, which is of about $\lambda /4,$ where $%
\lambda $ is the wavelength of our incident wave. This is an \emph{%
unexpected surprise}. From a purely angular spectrum argument, the spread of
backscatter angles for a fixed frequency would suggest a resolution of half
a wavelength. However, there has been previous evidence reported that using
a nonlinear inverse scattering algorithm for which strong or multiple
scattering occurs, that some degree of super resolution (i.e., beyond the
ideal diffraction limit of half a wavelength) can occur, see, e.g., \cite%
{Simonetti:APL2006}. This should be studied further.

There were also three data sets (\#23 in Table \ref{ta:2} and \#6, 20 in
Table \ref{ta:3}) in which one strong target and one weak target were buried
simultaneously. In all these three cases we have accurately imaged the
stronger target. However, we missed the weaker ones. The reason why we
missed the weak targets was due to the fact that their burial depths were
larger than 5 cm, which is our limiting depth for weak targets. %
%

Also of interest are three cases of \textit{heterogeneous} targets (\#11, 12
in Table \ref{ta:2} and \#18 in Table \ref{ta:3}), since explosive
devices are heterogeneous sometimes. We successfully estimated the average
refractive index of the geode, which consists of two different layers, in
data set \#12. For the wooden doll containing a metallic block inside in
data set \#18 the computed dielectric constant is larger than that of the
wood but smaller than other metallic targets. It is smaller because the wood
covers the metal. Target \#11 was a wooden doll with randomly distributed
metal screws inside. In this case we observed a weak signal, rather than a
strong one from the metal. In fact, we observed a well known phenomenon of 
\emph{diffuse scattering}, which was described in \cite{SGS:JOSA2002}. This
can be explained since the metal screws were randomly oriented and represent
a conducting very rough surface to the incident microwave pulse. Multiple
scattering combined with the penetration of the microwaves into the gaps
between the screws strongly attenuates the incident wave and little scatters
back to contribute to a measurable signal.

The reconstruction of shapes of the targets is illustrated in Figures \ref%
{fig:shape0}--\ref{fig:shape3}. For a better visualization, we show in these
figures the targets in the box with sizes $0.4\times 0.4\times 0.24$, which
is smaller in $x,y$ directions than our computational
domain $\Omega $. To show the shapes of the targets, we computed a truncated
coefficient $\epsilon _{t}(x,y,z)$ as follows: 
\begin{equation*}
\epsilon _{t}(x,y,z)=%
\begin{cases}
\epsilon (x,y,z)\text{ if }(x,y)\in \Gamma _{T}\text{ and }\epsilon (\mathbf{x})>\gamma \epsilon _{max} \\ 
\epsilon (\text{sand})\text{ otherwise},%
\end{cases}%
\end{equation*}%
for strong targets, with $\epsilon _{max}$ is the maximum of $\epsilon (%
\mathbf{x)}$. For weak targets, $\epsilon _{t}(x,y,z)$ is given by 
\begin{equation*}
\epsilon _{t}(x,y,z)=%
\begin{cases}
\epsilon (x,y,z)\text{ if }(x,y)\in \Gamma _{T}\text{ and }\epsilon (\mathbf{x})< \gamma \epsilon _{min} \\ 
\epsilon (\text{sand})\text{ otherwise},%
\end{cases}%
\end{equation*}%
where $\epsilon _{min}$ is the minimum value of $\epsilon (\mathbf{x)}$.
Recall that $\Gamma _{T}$ is the estimated $xy$ cross-section of the target,
see section \ref{subsec:twotest}. In this paper, the truncation parameter $%
\gamma $ was chosen $\gamma =0.7$ for Test 1 and $\gamma =0.6$ for Test 2.

Figure \ref{fig:shape0} depicts the reconstruction of target \#2 (a metallic
ball buried at 3 cm depth). We can see that the shape is quite well
reconstructed in Test 2, especially the $xy$-cross section, since
the measured data was acquired in the $x,y$-plane. Figure \ref{fig:shape1}
shows the reconstruction of target \#3 (a plastic bottle filled with a clean
water). In this case, since the target was quite high (about 18 cm), the
incident wave was weak at the vertical ends of the bottle which made it
difficult to reconstruct the complete shape. However, we still can see in
both tests the stretch in the vertical direction following the shape of the
bottle. Figure \ref{fig:shape2} illustrates the reconstruction of target
\#10, a ceramic mug. This is a weak target. We can see that its shape is
less accurately reconstructed than ones of strong targets, since its signal
is weak. Finally, Figure \ref{fig:shape3} shows the reconstruction of the
most difficult case of data set \#25: the two targets at 1 cm separation.
This seems to be a super resolution case as we discussed above.

\section{Conclusions}

\label{sec:con}

We have demonstrated the performance of the globally convergent algorithm of 
\cite{B-K:2012} for twenty five (25) test objects buried inside a sand box,
which models imaging of subsurface objects. Since the signal from the sand
is mixed with the signal from the target, this case is much harder than our
previous results for targets placed in air \cite{BTKF:IP2013,TBKF:SISC2014}. 
Our results have shown that it is possible to image refractive
indices of non metallic and effective dielectric constants of metallic
buried objects using backscattering time dependent measurements associated
with only one incident wave. In general, the reconstruction results
presented here are quite accurate. In particular, we have shown that the
technique of \cite{B-K:2012} can image quite high target/background
contrasts in dielectric constants (Table \ref{ta:3}), which is usually hard
to achieve by locally convergent algorithms. There are some cases in which
weak targets were missed when they were in placed together with strong
targets. This was due to the fact that their signals were too weak
to be detected since they were buried at deep depths exceeding our limiting
depth of 5 cm. This made us unable to extract their signals.

To improve the accuracy of reconstruction of shapes of targets, we
plan to use the adaptivity technique on the second stage of our two-stage
numerical procedure, see for results for targets in air \cite{BTKM:IP2014}.
We also plan to generalize the current globally convergent inversion algorithm to the case of the frequency
domain data in which the extraction of the target's signal will no longer be
needed. This might lead to a lesser number of missed weak targets.

\section*{Acknowledgment}

This research was supported by US Army Research Laboratory and US Army
Research Office grants W911NF-11-1-0325 and W911NF-11-1-0399, the Swedish
Research Council, the Swedish Foundation for Strategic Research (SSF) through the
Gothenburg Mathematical Modelling Centre (GMMC) and by the Swedish
Institute, Visby Program.

The authors are grateful to Mr.~Steven Kitchin for his excellent work on
data collection.


\end{document}